# Revisiting QUAD's ambition in the Indo-Pacific leveraging Space and Cyber Domain

Author- Mandeep Singh Rai
Affiliation- The Fletchers School of Law and Diplomacy, Tufts University.
Corresponding author email address- Mandeep.Rai@tufts.edu, msrai1980@gmail.com

## Abstract

In the backdrop of growing power transitions and rise of strategic high tide in the world, the Indo Pacific Region (IPR) has emanated as an area which is likely to witness increased development & enhanced security cooperation through militarization. With China trying to be at the seat of the Global leadership, US & its allies in the Indo pacific are aiming at working together, finding mutually beneficial areas of functional cooperation and redefining the canvas of security. This purpose of this paper is to analyze an informal alliance, *Quadrilateral Security dialogue* (QUAD) in its present form with the geostrategic landscape in Indo pacific and present recommendations to strengthen collaboration & response capability. To advance the main arguments, this paper project proceeds as follows.

In Chapter 1, we furnish a big picture on the changing geopolitical and geoeconomics scenario due to Chinese expansionary attitude in Indo Pacific, which make the region significant and motivates for an inquiry into its future course of action & alignment. Chapter 2, collates the arguments about the existing geopolitical shift in the international relations which pave the way for formal and informal alliances by comparing their framework. Next, we examine the interests and motives of QUAD countries in Indo Pacific with reference to growing Chinese aggressive strides in the region. We probe Chinese strategy in Indo Pacific & its manifestation as a coercive power leveraging economic investment for projection in the maritime domain. In Chapter 3, through our literature survey, we show that the existing work does not highlight specific scope for collaboration by the QUAD countries especially in the Cyber and Space domain and there is a need to focus on distinct use cases in both domains. Against this back drop, Chapter 4, covers deliberation on the present work and provides specific policy recommendations for the alliance. Lastly, in Chapter 5, we conclude with our summary by visualizing the way forward for the QUAD alliance based on evolving geopolitical and security landscape.

# Table of Contents

## 1. Introduction & Motivation

Global world order and geopolitics have witnessed major shocks in recent years in the backdrop of COVID 19 pandemic, the Taliban 2.0 episode and the recent Ukraine crisis. A major shift in international relations and norms is taking place with several countries opting for alliances adapting to changed security threats. The power parity in the present world order has begun to drift from the Atlantic to the east, making Indo pacific as one of the hotspots of geopolitical rivalry which basically paves way for the conceptualization and concretization of the Indo pacific[1]. The year 2021 witnessed a substantial awaking in the Indo pacific with major policy, framework and posture adaptations by countries within & outside the region. Globally, the dynamic geopolitical landscape is making countries choose sides and forge partnerships to contain the threat posed by growing rivalry between the US, Russia and China. Many new informal and formal partnerships are being initiated in the geostrategic space due to hardening power rivalry and competition as countries assess emerging threats, challenges and opportunities in the region.[2]

The Indo pacific region is a conflux of Indian Ocean and the Pacific ocean which serves as a nest of major rising powers. China, India and ASEAN have the world's largest population & GDP share globally, share world's busiest trade routes, and most importantly powerful military forces.

**Figure 1:** Map depicting the Indo Pacific region.

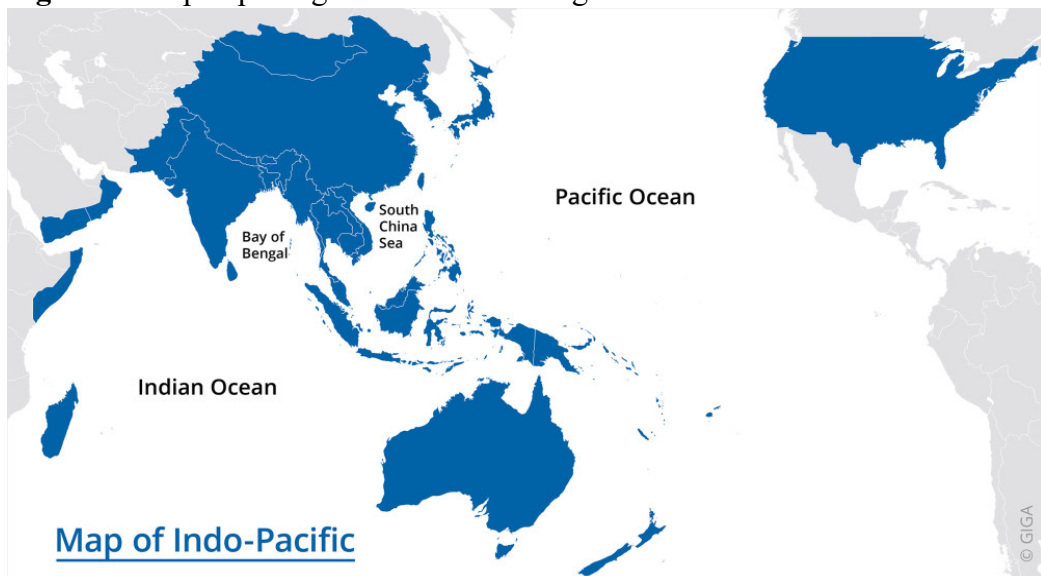

**Source** : David Camroux & Christopher Jaffrelot, "*The Concept of the Indo Pacific in the Geostrategic Discourse*".(https://www.sciencespo.fr/ceri/en/content/lindo-pacifique-quels-contours-quels-enjeux-0)

China, in the past decade has seen an exponential rise in its economic and military spheres. When seen in the light of Belt and Road Initiative (BRI), the sphere of influence looks limitless as the project not only focuses on economic leaps but strategic aspects as well.[3]

---

[1] Soumyodeep Deb and Nathan Wilson, "*The Coming of QUAD and the Balance of Power in the Indo-Pacific,*" *Journal of Indo-Pacific Affairs*, December 13, 2021 assessed on June 10 2022 at https://www.airuniversity.af.edu/JIPA/Display/Article/2870653/the-coming-of-QUAD-and-the-balance-of-power-in-the-indo-pacific/#sdendnote1sym

[2] Girish Luthra, "The Indo-Pacific region, maritime focus, and ocean governance", January 08, 2022, *Observer Research Foundation* assessed on June 10,2022 at https://www.orfonline.org/expert-speak/the-indo-pacific-region-maritime-focus-and-ocean-governance/

[3] Dr Aparaajita Pandey, "*The evolving dynamics between China and the QUAD*", May 24,2022, *Financial Express*, assessed on 01 July 2022 at https://www.financialexpress.com/defence/the-evolving-dynamics-between-china-and-the-QUAD/2536283/

the economic growth in the Indo Pacific continues to shift the strategic landscape, it is accelerating tectonic changes to balance economic and strategic weight both in the US and EU Indo pacific region.[4] (Ref Figure 2)

**Figure 2:** Map depicting BRI, US Indo pacific & EU Indo-pacific region.

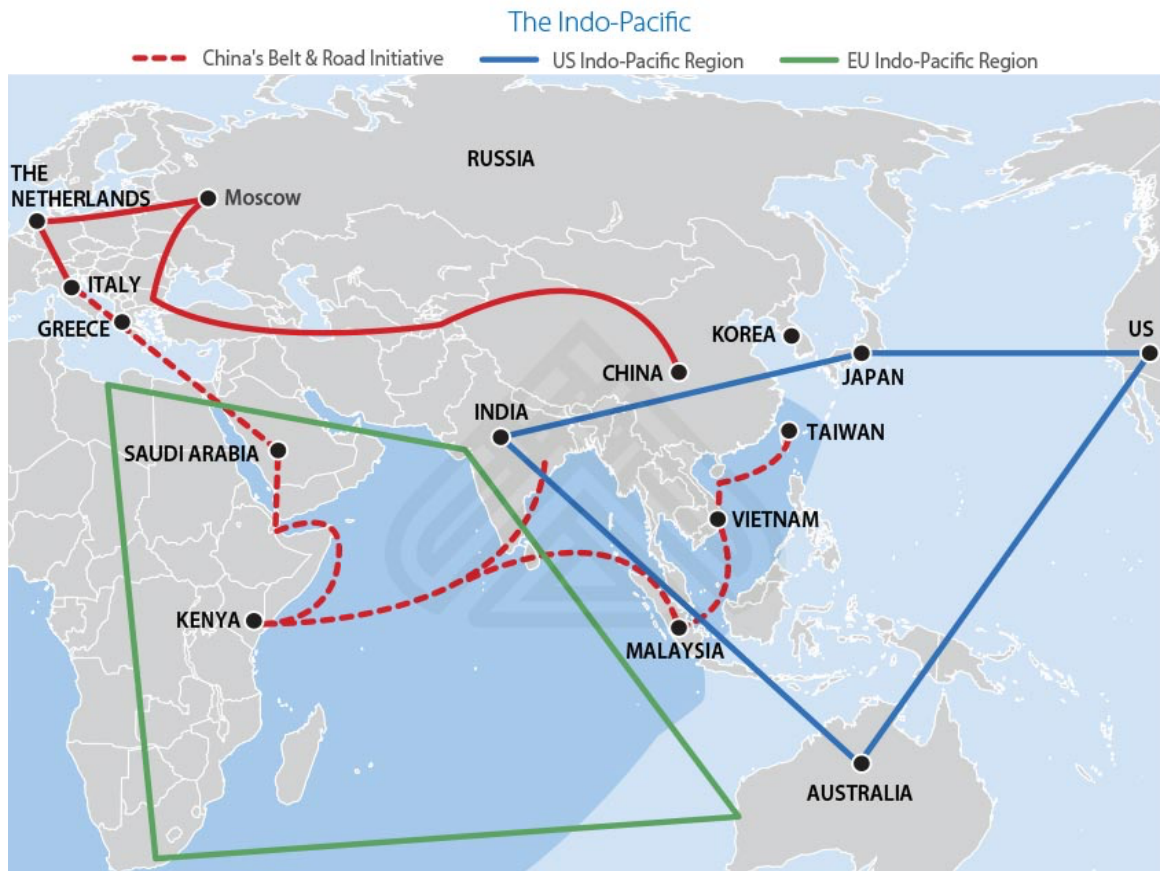

**Source:** Andrew Wheeler, "Has China's Belt and Road Initiative Shifted the Geo-Political Regional Debate from APAC to the Indo-Pacific? "*Silk Road Briefing*, (https://www.silkroadbriefing.com/news/2020/11/04/has-chinas-belt-and-road-initiative-shifted-the-geo-political-regional-debate-from-apac-to-the-indo-pacific/)

With growing influence of China particularly in South East Asia, maritime and border disputes will perpetuate to create friction. It is imperative that the long term interest of the regional powers in military and economic domains converge to stabilize the balance of the regional order and desire to counterbalance China gets adequate focus. [5] Figure 3 provides a snapshot of the incidents specifically in South China Sea (SCS) from local and international news resources from 2012-2019 and it is found that 52% of the reported incidents were from Chinese vessels.[6]

---

[4] Australian Government, (2017) "*Power shifts in the Indo-Pacific*" 2017 Foreign Policy White paper, assessed on 10 July 2022 at https://www.dfat.gov.au/sites/default/files/minisite/static/4ca0813c-585e-4fe1-86eb-de665e65001a/fpwhitepaper/foreign-policy-white-paper/chapter-two-contested-world/power-shifts-indo-pacific.html
[5] Ibid,4.
[6] "*South China Sea Data initiative"* assessed on July 10, 2022 at *https://www.scsdi.org/data.*

**Figure 3:** Percentage of incidents involving country vessels from 2012 to 2019.

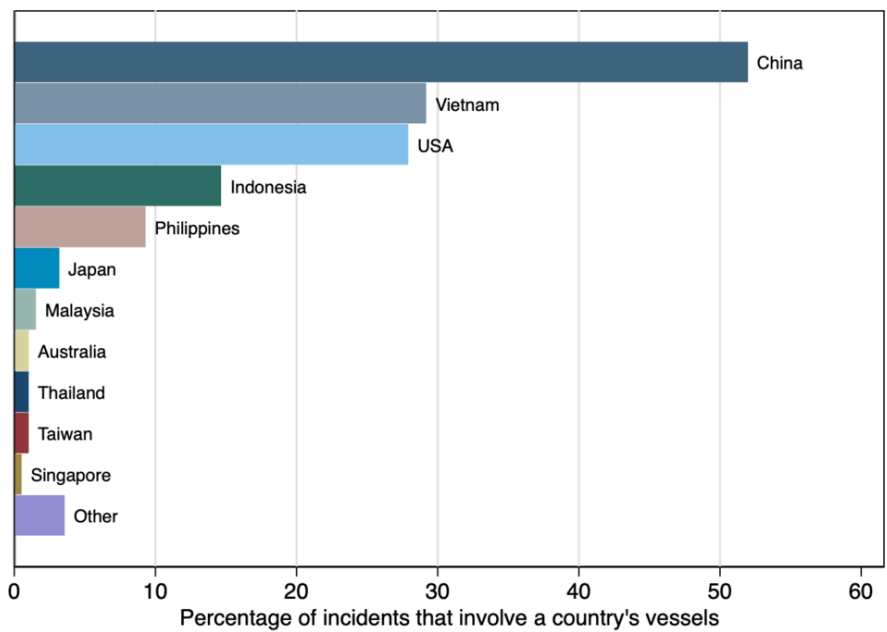

**Source**: *South China Sea Data initiative (https://www.scsdi.org/data)*

The *Quadrilateral Security Dialogue* colloquially called QUAD, is an informal alliance between US, India, Japan and Australia which has been revived due to common threat perception by QUAD countries to counter China's rising influence in the Indo Pacific region. What started as an adhoc arrangement of like-minded countries during the 2004 Tsunami has now established itself as a critical pilaster in the Indo Pacific region with a more assertive and shifting tone from the past. At present, QUAD has been raised up to a leader level dialogue which has developed a refurbished working group system for cooperation significantly broadening and deepening their vision on agendas in the field of climate change, vaccines, critical & emerging technologies, infrastructure, cyber and space.[7]

The current stature of QUAD raises certain questions on its future course of action & alignment with the changing geopolitical and geoeconomics scenario.

a) How well the informality of the QUAD alliance will pay expected dividends and set the course for containing assertive China?

b) Will the working group structure deliver concrete results in the fields they have been conceived? Is there a flexibility within the alliance that can transform it into a formal alliance at the time of crisis?

c) Are the areas of cooperation overcoming strategic challenges & whether the countries will be able to accrue benefits in terms of strategic and security needs?

d) Have specific areas in the agenda been addressed where collaboration can benefit not only the QUAD countries but other countries in the Indo Pacific?

---

[7] Garima Mohan & Kristi Govella, "*The Future of the QUAD and the Emerging Architecture in the Indo-Pacific*", June 2022, *The German Marshal Fund of the United States, assessed on 10 July 2022* at https://www.gmfus.org/sites/default/files/202206/The%20Future%20of%20the%20QUAD%20and%20the%20 Emerging%20Architecture%20in%20the%20Indo-Pacific.pdf

Above questions are thought provoking & worth considering to visualize the future construct of the QUAD alliance. This paper aims to address specific areas which needs collaboration by the QUAD countries within the working group system for a more constructive and efficacious partnership. In the next sections, we proceed by collating and summarizing the arguments on the existing geopolitical shift in the international relations which pave the way for formal and informal alliances and in the context, consider emergence of QUAD. We examine the interests and motives of QUAD countries for collaboration with common interests and how they envision the Indo Pacific region as a pivot for strengthening ties. Next, we provide a forecast on how China will expand in the region in the garb of economic investment through its ambitious BRI and use the infrastructure developed for projection in the maritime domain conforming to bolster its strategy to evolve as a global leader. We then take two crucial fields of Space and Cyber security where through our literature survey, we show that at present the existing work does not provide key areas for collaboration by the QUAD countries and needs focus on specific segments in both domains. Against this back drop, we predict the future manifestation of QUAD as an alliance and provide specific recommendations for its way forward in the coming years.

## 2. Background and basics

### 2.1 IGOs as formal and Informal alliances

The vacuum created by the shifting power balance in the present geopolitical world has led countries to collaborate multilaterally through formal & informal alliances to protect their geostrategic, geopolitical and economic interests. Scholars of International relations (IR) scholars have observed IGOs primarily as agents of states and have concentrated on their structural characteristics, governing processes, and programs.[8] IGOs are perceived to be actors with whom, the members persuade states to act by coordinating their efforts in providing them diplomatic skills for securing agreements to ensure effectiveness of programs.[9] The functions of IGOs have been identified as '*informational*', in which gathering, analysing and dissemination of information is undertaken or '*normative and rule creation*' which involves supervising & allocating resources for operational mechanism. IGO's prove to be one of the means which the emerging, regional and small powers use for power transitions to acclimate them for own political and economic benefits. For example, NATO during its transformation in 1990's was portrayed from a well-defined military alliance to an arrangement which offered "collective security" and a "promoter of democracy" in order to expand its membership.[10] Figure 4 provides an illustration of how a formal alliance NATO, has grown over the years from a dozen countries to 30 as of 2020.

**Figure 4:** Growth of NATO from 1949-2020

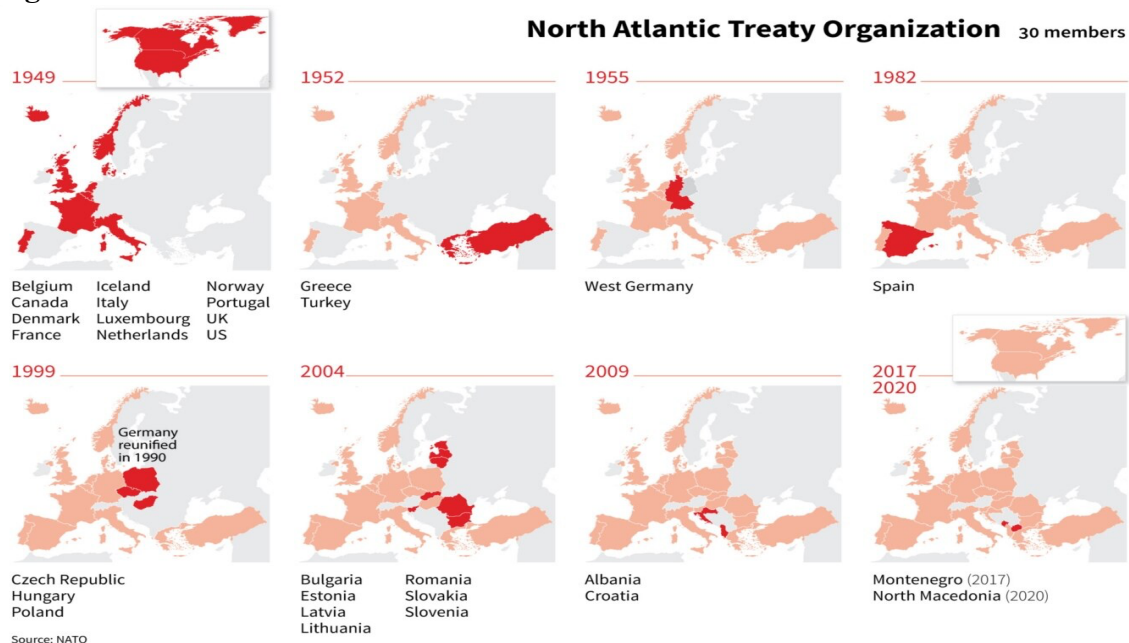

**Source**: The Growth of NATO, February 07, 2022, *VoaNews*, (https://www.voanews.com/a/the-growth-of-nato/6431174.html)

Nation states within themselves have asymmetric advantages of governance apparatus for mobilizing its population but lack collective mobilization across sovereign borders.[11] International organizations provide them with a platform for collective mobilization to

---

[8] Karns, Margaret P., et al. International Organizations : The Politics and Processes of Global Governance, Lynne Rienner Publishers, 2015
[9] Ibid.
[10] "*The Debate on NATO Enlargement*", Hearings before the committee on foreign relations United States Senate, October 7,1997, U.S. Government printing office, assessed on 24 June 2022 at https://www.govinfo.gov/content/pkg/CHRG-105shrg46832/html/CHRG-105shrg46832.htm
[11] Ibid.

engage in a joint action.[12] The creation of these formal IGO's reflect a balanced equation between the powerful countries and their weak counter parts. The powerful states accept the institutional norms to set the rules and the weak states accrue the benefits out of the rule based mechanism to counter the power driven governance by other hegemons.[13] Vabulas and Snidel provide a theoretical analysis of IGO's. The author's analyse a spectrum of arrangements to highlight significant variation in formalization of the international institutions and proclaim why the choices are driven to informal arrangements over a formal.[14] IIGOs (Informal Intergovernmental Organizations) provide an increased flexibility in comparison to FIGO's (Formal Intergovernmental Organizations) which have a binding commitment of states. They maintain state autonomy with minimal bureaucratic costs and have a lower transaction cost for implementation. As far as policy implications are concerned, IIGO's are apt choices to mediate power shifts as the nature of flexibility provides a forum for gradual adaptation before reaping the long term benefits.[15] If we track the growth of the IIGO's (which include one or both US and China since 1940), it is assessed that IIGO's have risen dramatically after the China's power shift in 1990's by approximately 68 percent in comparison to FIGO's which have remained dormant. [16] (Refer Figure 5)

**Figure 5:** Shift to IIGOs during China power shift.

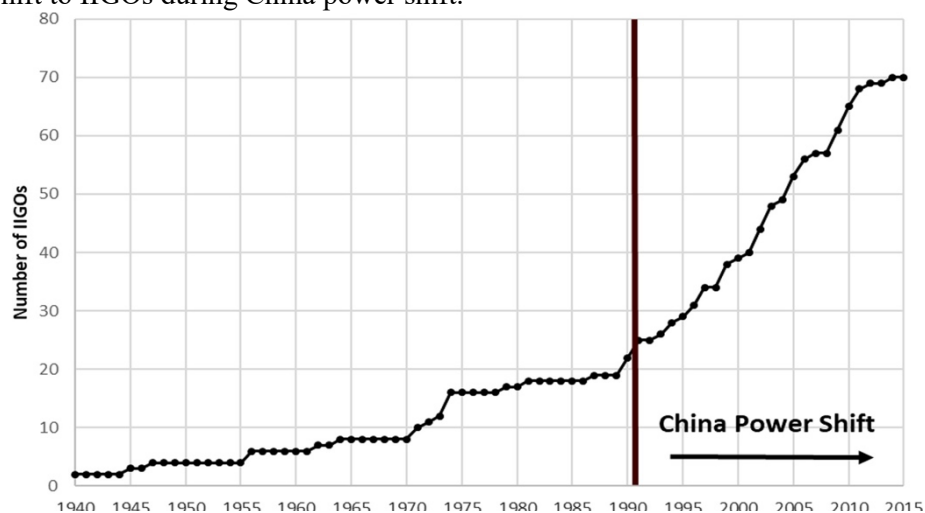

**Source**: Vabulas, F. and Snidal, D. (2020), *Informal IGOs as Mediators of Power Shifts*. Glob Policy, 11: 40-50., fig.1.

A prime reason for countries to opt for IIGOs is the amount of flexibility which the IIGO's provide to adapt to dynamic changes & respond spontaneously without enforcing a nation state into an institutional distribution of power. The mode of institutional change which has driven the IIGO's post 1990's fall in the category of '*system management*' of powershift category where collaboration of like-minded nations and hegemonic consensus favours for a deeper understanding over a common cause in contrast to *redistributive* and *integrative* strategies. [17]

---

[12] Norrin M. Ripsman " A Neoclassical Realist Explanation of International Institutions" in Anders Wivel and T. V. Paul, eds., International Institutions and Power Politics: Bridging the Divide (Washington, DC: Georgetown University Press, 2019) pp. 41-52
[13] Ibid.
[14] Vabulas, Felicity. (2013). Organization without Delegation: Informal Intergovernmental Organizations (IIGOs) and the Spectrum of Intergovernmental Arrangements. The Review of International Organizations. 8. 10.1007/s11558-012-9161-x.
[15] Vabulas, F. and Snidal, D. (2020), Informal IGOs as Mediators of Power Shifts. Glob Policy, 11: 40-50.
[16] Ibid.
[17] Ibid,15.

## 2.2 Characterisation of QUAD as an informal alliance.

The QUAD displays many attributes of a highly informal IGO and naming it as a security dialogue forges it to be a dialogue, which is fallacious and lacks substance.[18] The informality of QUAD alliance in all proportion, suits its members as it is mutually constitutive & reinforcing. In addition, it places minimal obligations on member states over converged interests. Its colloquial nature allows the member states to signal different messages as opportunities in the shadow of strong partners to convey hard-line stands. As an example, Japan sees the hard-line stand of US against China as an opportunity not to issue direct statements on its own rather leverage the necessary space created by Washington's language to its own advantage on contentious issues with China.[19] The conglomeration proves itself as dual purpose tool to counter balance Chinese hegemonism and as a diplomatic hedging mechanism. Its ambiguous form provides flexibility in its form, allowing it to be augmented in response to any negative behaviour by China, or taper down if it behaves well. Its submerged position, intrigued structure and pliability puts the liability back on China court.[20]

## 2.3. Growing interests of QUAD nations in the Indo pacific

In the 21$^{st}$ century, Indo pacific region has been recognised as one of the most vital to prosperity and security in the global context. It attracts key stake holders to bring their interests on a common platform to gain strategic value. The region faces aggravated challenges through coercive and aggressive moves of China in the economic, military and technological domain. This section analyses the strategic interests of QUAD countries in the Indo Pacific, reflecting importance of building collective capacity by nation states as part of regional alliances to advance their broad strategic interests.

### 2.3.1 United States Indo Pacific Strategy

US has a presence of maximum "boots on ground" in the Indo Pacific region than any other offshore base. It is a vital source of Foreign Direct Investment (FDI) and a source of approximately 3 million jobs which creates a manoeuvre space of potential importance for the US in the future. [21] US recognises Indo Pacific as part of its vital strategy for its consistent role and commitment in the region as it transforms to become the center of gravity in the years to come.[22] Some of the challenges posed to the status of US as a dominant power are due to Beijing's continued military, economic and diplomatic power plays in Asia.[23] China's adventurism in the East and South China Sea and more so its posturing with the key US allies Taiwan, Japan, Australia and India is a major cause of concern for the US dominance.[24] In the month of February 2022, US released its Indo Pacific Strategy (IPS) which lays out the narrative of its interest in the region & in support of a collective response against the Chinese growing influence. The themes presented in the policy as "Driving Indo Pacific prosperity" and "Bolster Indo pacific security" presents an innovative framework for close economic integration and in aid to regional peace, security and stability.[25] The IPS makes a 'qualifier response' to the open competition between US & China, intended as a sign of a barge in style from the previous administration. However, the release of the strategy before the revised "National Security Strategy" & "National Defence Strategy" puts a factor of uncertainty in the assertive tone of the US.[26] In the Obama administration, TPP (Trans Pacific partnership) was thought as a tool to rewrite the rules of engagement with China, however the failure of TPP meant that US should come up with a revised "Integrated Deterrence Policy". US focuses the strategic competition against China by leveraging key military advantages, promote democratic values, invest in niche technologies and revise vital security & economic partnerships.

---

[18] Cannon, B.J., Rossiter, A. Locating the QUAD: informality, institutional flexibility, and future alignment in the Indo-Pacific. *Int Polit* (2022). https://doi.org/10.1057/s41311-022-00383-y
[19] Ibid.
[20] Graham, Euan & Pan, Chengxin & Hall, Ian & Kersten, Rikki & Zala, Benjamin & Percy, Sarah & Carr, Andrew & Australian National University. Strategic and Defence Studies Centre, issuing body.
(2018). *Debating the QUAD* Retrieved July 3, 2022, from http://nla.gov.au/nla.obj-1029574453
[21] *Indo pacific Strategy of the United States*, February 2022, *The White House,* assessed on 14 June 2022 at https://www.whitehouse.gov/wp-content/uploads/2022/02/U.S.-Indo-Pacific-Strategy.pdf
[22] Ibid,20.
[23] "Reassessing China's Threat to U.S. Power Abroad", February 19, 2021,usnews, assessed on 10 July 2022 at https://www.usnews.com/news/best-countries/articles/2021-02-19/does-china-really-threaten-us-power-abroad
[24] Jagannath Panda and Akriti Vasudeva, "*US- China Competition and Washington's case for 'QUAD plus'*" *The National Interest*, ,September, 28, 2022 assessed on 14 June 202 at https://www.stimson.org/2020/u-s-china-competition-and-washingtons-case-for-QUAD-plus/

Washington aspires toward a Free & Open Indo-Pacific (FOIP), defined on the pillars of respect for sovereignty, independence of nations, greater transparency, peaceful resolution of disputes, free and fair trade, adherence to international law, and good governance.[27] It pledges support for deepening bilateral & multilateral partnerships, regional connectivity, trade & investment, and emphasizes cooperation with regional partners through "Indo Pacific Economic Framework", to identify new propositions to trade which meet high labour and environmental standards.[28] The *'integrated deterrence'* which is supposedly the corner stone of the US approach, will potentially be challenging to integrate efforts across all spectrum of conflicts to counter coercion and reinforce deterrence.[29] Against the increase sophistication and aggressive behaviour of PLAN and Russian federation, US has conceptualized the strategy of "Advantage at Sea"[30] to prioritize controlling of seas, increase its forward deployment and realign warfighting organizations. At present, approximately 60 percent of the US forces are in the Indo Pacific region.[31] There is a massive transformation of the Marine Corps being completed to generate greater expeditionary power with enhanced sea denial and control capabilities.[32]

**2.3.2 Australia and the Indo Pacific**

Australia has emanated as a key strategic player in the Indo-Pacific region in the midst of strategic shifts. China endures to be Australia's largest trading partner[33] and both countries have displayed growing concerns over each other's stance during multilateral forums. Its geographic location bordering Indian ocean to the west, Pacific Ocean in the east and the ASEAN states in the north makes it's a centre piece in the Indo Pacific.

---

[25] Philip Shetler-Jones, "*The strange case of America's Indo Pacific Strategies*", March 14 2022, *Chatham House*, The Royal Institute of International Affairs assessed on 16 June 2022 at
https://www.chathamhouse.org/2022/03/strange-case-americas-indo-pacific-strategies
[26] "*Indo-Pacific Strategy Report: Preparedness, Partnerships, and Promoting a Networked Region*", U.S. Department of Defense June 1, 2019, assessed on 15 June 2022
at https://media.defense.gov/2019/Jul/01/2002152311/-1/-1/1/department-of-defense-indo-pacific-strategy-report-2019.pdf..
[27] Ibid
[28] Carla freeman, Daniel Markey and Vikram J. Singh, "*A closer look at Biden's Indo Pacific Strategy*" March 7, 2022 *United States Institute of Peace* assessed on 17 June 2022 at
https://www.usip.org/publications/2022/03/closer-look-bidens-indo-pacific-strategy
[29] Ibid, 26.
[30] "*Advantage at Sea- Prevailing with Integrated All-Domain Naval power*", December 2020, assessed on June 17 2022 at https://media.defense.gov/2020/Dec/16/2002553074/-1/-1/0/TRISERVICESTRATEGY.PDF
[31] Krishn Kaushik,"*60% Navy forces in Indo-Pacific region now: US Navy chief*", October 13, 2021, *Indian Express,* assessed on 10 July 2022 at https://indianexpress.com/article/india/60-per-cent-navy-forces-in-indo-pacific-region-now-us-navy-chief-7568984/
[32] Ibid, 36.
[33] Weizhen tan, "*Australia's exports to China are jumping despite their trade fight*", October 27, 2021, *CNBC*, assessed on 03 July 2022 at https://www.cnbc.com/2021/10/28/australias-exports-to-china-are-jumping-despite-their-trade-fight.html

The Australian perspective of the balance of economic opportunities & strategic challenges in the Indo Pacific region began to see new light, with latter proving to be of greater significance during successive defence and foreign policy white papers namely: The 2016 Defence White Paper (DWP), The 2017 Foreign Policy White Paper (FPWP), and the latest, The 2020 Defence Strategic Update (DSU).[34] The DWP which act as vision document for 20 years focuses on six key drivers shaping Canberra's strategic environment: 1) China & US cooperation and competition 2) challenges to the rules based order 3) terrorism 4) state fragility 5) military modernisation and 6) new complex & nongeographic threats, including cyber-attacks.[35] Australia aims to strengthen ties with US & other nations to manage uncertain security environment and commits an increased defence spending over a eleven year period.[36] Canberra's Defence Strategic Update-2020 provided the guidance towards this change "*to ensure Australia is able to deploy military power to shape its environment, deter actions against its interests and, when required, respond with military force*" focusing on Indo-Pacific.[37]

In December 2020, Australia voiced for an independent investigation on the origins of COVID-19 pandemic[38], which was accused by Beijing and led to targeting of Australian imports particularly, barley, beef, wine and coal. Canberra has discerned that its interests in the Indo-Pacific region are unprotected & vulnerable to economic coercion, disruptions and grey zone attacks by adversaries.[39] As mass proportion of its trade passes through sea lanes of South East Asia[40] and Indian Ocean, Australia at present is intensifying its interests in the Indo Pacific region by conceptualizing a robust strategy to prioritize its potent capabilities. It has been an alliance partner with the US for past 70 years since the inception of 1951 ANZUS (Australia, New Zealand and United States) Security Treaty[41]. The AUKUS (Australia-UK-US) treaty which came into effect just before the September QUAD Summit[42], has further strengthened the alliance as they reviewed the progress of the partnership towards nuclear stewardship equipping Australian Navy a heft in the pacific.[43] The AUKUS leaders during the meeting on 5 April 2022 reaffirmed their resolve & commitment to a FOIP and to an international system having respect for human rights, rule of law, and peaceful resolution of disputes free from coercion.[44] The implementation of the AUKUS will arm Australia with conventionally armed nuclear submarine capability and provide optimal pathway to establish joint advanced military capabilities by in developing undersea capabilities, accelerate investment in niche technologies and advanced cyber capabilities & electronic warfare capabilities.[45]. Australia has been challenging the assertive and coercive strides of Beijing since 2016 when it became the first country to recognize the 2016 Permanent Court of Arbitration decision against China to abide by the ruling.[46] Expressing its vital concerns on the national security, Australia was amongst the first country to ban Huawei from participating in the rolling out of the 5G network[47]. Canberra acknowledges that the Chinese BRI has an attractiveness that makes the Pacific Island nations vulnerable to falling prey to 'debt trap diplomacy'.[48] It asserts to expand its efforts to portray itself as a reliable partner in the Pacific Island nations and has stressed basing its relations on openness, respect, and equality.[49]

---

[34] Matthew Parry; Graphics: Lucille Killmayer., 2022 "*Australia's strategic view of the Indo Pacific*", EPRS: European Parliamentary Research Service, assessed on 17 June 2022 at
https://www.europarl.europa.eu/RegData/etudes/BRIE/2022/698917/EPRS_BRI(2022)698917_EN.pdf
[35] Ibid,20.
[36] Australian Government (2016), *2016 Defence White Paper* (2016), Department of defence
[37] "*2020 Defence Strategic Update*", Department of Defence, Australian Government.
[38] Daniel Hurst, ""*Australia insists WHO inquiry into COVID origin must be robust ,despite China tensions*", December 28, 2020 , *The Guardian*, assessed on 24 June 2022 at
https://www.theguardian.com/world/2020/dec/29/australia-insists-who-inquiry-into-covid-origin-must-be-robust-despite-china-tensions
[39] Singh, R., 2022. "*Australia's Strategic Imperatives In Indo-Pacific: Opportunities For India*", Manohar Parrikar Institute for Defence Studies and Analyses, assessed on 14 June 2022 at
https://www.idsa.in/issuebrief/australias-strategic-imperatives-in-indo-pacific-rpsingh-230222
[40] Sean Andrews and Alastair Cooper, "*An Australian vision of the Indo-Pacific – Through a Strategic and Maritime Lens*", Sea power, review and Analysis, Issue1, 2019,Royal Australian Navy, assessed on 12 June 22 at https://www.navy.gov.au/sites/default/files/documents/Australian_vision_of_Indo-Pacific_Through_Strategic_and_Maritime_Lens_0.pdf
[41] Patricia O'Brien, "*The ANZUS Treaty at 70*", August 25, 2021, *The Diplomat,* assessed on 24 June 2022 at https://thediplomat.com/2021/08/the-anzus-treaty-at-70/
[42] Akshobh Giridharadas, "*Why the AUKUS help the QUAD*", October 21, 2021, Observer Research Foundation assessed on June 15,2022 at https://www.orfonline.org/expert-speak/why-the-aukus-helps-the-QUAD/
[43] Krishna Kaushik, "*Explained: The AUKUS agreement to equip Australia with n-subs, and why it has upset France*", September 18, 2021, *The Indian Express,* assessed on 24 June 2022 at
https://indianexpress.com/article/explained/aukus-agreement-to-equip-australia-with-nuclear-submarines-7513013/

### 2.3.3 Japan and the Indo pacific

Historically a trading nation, Japan has remained concerned about its maritime focus even after being demilitarized after the World War II. Post the world war, Japan has emerged as a leading commercial industrialized country in ship building and fisheries under the shadowed maritime security through the US dominance. The growing concern with the Chinese increased footsteps in the maritime domain has led Japan to rethink its perceptions in the Indo pacific region.[50] Late Japanese Prime minister Shinzo Abe during his tenure gave an emphasis toward increased partnership in the Indo Pacific and referred to Japanese pillars of diplomacy symbolizing the need for "Arc of freedom and Prosperity"[51], "Confluence of two seas" [52] and "Asia's democratic Security Diamond"[53]. In 2014, Japan revised its policy on the long standing policy of armed exports which prevented Japanese industry to export weapons and related technology thereby agreeing to export services in surveillance, rescue, warning, transport and minesweeping.[54] In 2022, in order to strengthen its security cooperation, Japan allowed export of fighter jets, missiles and weapons to 12 nations (QUAD countries included) with an aim to bolster deterrence capability and counter its outlandish claims.[55] The core element of the Japan's IPS is centred around economic sectors namely energy and infrastructure. The energy sector prompted a policy decision for Japan to focus on becoming a key supplier of the LNG thereby reducing the trade imbalance with US and providing regional customers an alternative to China. Being a pioneer in building high tech infrastructure, Japan acquired a powerful edge over strengthening infrastructure governance by maximizing infrastructures' positive impact with economic efficiency and building resilience against natural disasters.[56] The Japanese Self Defence forces have been little effective due to less political support from the leaders. The little interoperability within its branches and with the US made a strong reason for Japan to have bilateral ties with the nations to fix it deficiencies and garner external support. [57]

---

[44] *Factsheet: Implementation of the Australia-United Kingdom-United States partnership (AUKUS),* The White House, April 05, 2022 assessed on 16 June 2021 at https://www.whitehouse.gov/briefing-room/statements-releases/2022/04/05/fact-sheet-implementation-of-the-australia-united-kingdom-united-states-partnership-aukus/
[45] Ibid.
[46] Julie Bishop, "*Australia Supports Peaceful Dispute Resolution in the South China Sea,*" Australian Government Department of Foreign Affairs and Trade, 2016.
[47] Jonathan Pearlman, "*Australia Bars Huawei from 5G Tender in Move Likely to Irk China,*" *Straits Times*, 23 August 2018, assessed on 12 June 2022 at https://www.straitstimes.com/asia/australianz/australia-bars-huawei-from-5g-tender-in-move-likely-to-irk-china
[48] Through the Debt Trap diplomacy, China is creating a strong position geopolitically by extending debt to countries to increase its political leverage.
[49] Rajah, Roland; Dayant, Alexandre; Pryke, Jonathan. 2019. *Ocean of debt? Belt and Road and debt diplomacy in the Pacific*. © Lowy Institute For International Policy. http://hdl.handle.net/11540/11721.
[50] Cleo Paskal, "*Indo pacific strategies, perceptions and partnerships*", March 23, 2021, © *Chatham House*, The Royal Institute of International Affairs, 2022 assessed on 22 June 2022 at
https://www.chathamhouse.org/2021/03/indo-pacific-strategies-perceptions-and-partnerships/07-japan-and-indo-pacific
[51] The "*Arc of freedom and Prosperity*" is a pillar of Japanese diplomacy for attainment of universal values. The Arc starts from Northern Europe, traverses through the Baltic states, Central and South Europe, Central Asia, Middle East, Indian subcontinent, reaches the Southeast Asia and culminating at Northeast Asia.
[52] "*Confluence of the Two Seas*" Speech by H.E.Mr. Shinzo Abe, Prime Minister of Japan
at the Parliament of the Republic of India, Ministry of Foreign Affairs of Japan, August 22, 2007 assessed on 02 July 2022 at https://www.mofa.go.jp/region/asia-paci/pmv0708/speech-2.html
[53] "*Arc of freedom and Prosperity*" was mooted by Late President Shinzo Abe of Japan as a strategy where India, Australia Japan & US form a diamond from Indian Ocean to Western pacific.
[54] "*The Three Principles on Transfer of Defense Equipment and Technology*" Japan's Security Policy, Ministry of Foreign Affairs of Japan, April 6, 2016 assessed on 24 June 2022 at
https://www.mofa.go.jp/fp/nsp/page1we_000083.html

In the disputed Senkaku island of Japan, China has employed its Coast guard vessels and its deliberate activities have increased manifold (From 12 incidents in 2011 to 1,097 in 2019).[58] Japan has been expressing its strong concern on the Chinese ships violating the international law and each time Chinese coast guard ship intrude Japanese territorial waters, Japanese patrol vessels confront them and simultaneously a protest is lodged through the diplomatic channel. [59] Taking into consideration the strategic uncertainty over Japan's international coalition-building initiatives to create a new regional order, it has used 'tactical hedging' as a tool by taking a flexible approach in its ambiguous FOIP concept to gauge states responses, interpret their perspectives, and its strategic emphasis accordingly.[60] Japan has been intensifying its effort under the "Proactive Contribution to International Peace" in line with its national security strategy and has made an effort to "internationalize" Chinese behaviour by arguing that other nation states in the South China Sea are equally affected by similar challenges which Japan faces in the East China Sea.[61]

**2.3.4 India and the Indo pacific.**

India has been concerned about its strategic and military posture from a mainly continental perspective as it has been facing disputes with its largest neighbours, China and Pakistan. In the past, India has underutilized the maritime sphere for economic growth and development as well as for expanding options with Pakistan and China. As India is located in the heart of the Indian Ocean region and shares 3,488 Km land border with China[62], it presents both

---

[55] Ashish Dangwal, "*In A Massive Policy Change, Japan Permits Export Of Its Fighter Jets, Missiles To 12 Countries Including India – Reports*", May 28, 2022, *The EurAsian Times*, assessed on 26 June 2022 at https://eurasiantimes.com/japan-permits-export-its-fighter-jets-missiles-to-12-countries/

[56] Paul Kriss & Darwin Marcelo, "*Urban infrastructure in Japan: Lessons from infrastructure quality investment principles*", March 22,2021, *World bank Blogs*, assessed on June 24, 2022 at https://blogs.worldbank.org/ppps/urban-infrastructure-japan-lessons-infrastructure-quality-investment-principles

[57] Satoru Nagao, "*Oceanic Choices: India, Japan, and the Dragon's Fire: How does the QUAD Work?*", April 28, 2022, *Observer Research Foundation* assessed on June 6 ,2022 at https://www.orfonline.org/expert-speak/how-does-the-QUAD-work/#_edn2

[58] "*Trends in Chinese Government and Other vessels in the Waters Surrounding the Senkaku Islands, and Japan's response*", May 30, 2022" Ministry of Foreign Affairs of Japan. Assessed on 23 June 2022 at https://www.mofa.go.jp/region/page23e_000021.htma

[59] Ibid.

[60] Kei Koga, Japan's 'Indo-Pacific' question: countering China or shaping a new regional order?, International Affairs, Volume 96, Issue 1, January 2020, Pages 49–73

[61] Yuki Tatsumi, "*Is Japan Ready For The QUAD? Opportunities And Challenges For Tokyo In A Changing Indo-Pacific*", January 9, 2018, WAR ON THE ROCKS, Special Series-Southern (Dis)Comfort, assessed on 24 June 2022 at https://warontherocks.com/2018/01/japan-ready-QUAD-opportunities-challenges-tokyo-changing-indo-pacific/

[62] Ravi Buddhavarapu ,"*China is building a new bridge on a disputed Himalayan border, drawing ire from India*", May 24, 2022, *CNBC*, assessed on 21 June 2022 at https://www.cnbc.com/2022/05/25/china-is-building-a-bridge-on-a-disputed-himalayan-border-with-india.html

opportunities and challenges that are important in geopolitics and security dimension of Indo Pacific.[63] India adopts a well-defined view of the link between economic growth and prosperity with the help of global engagement in the Indo pacific. For contemporary India, development resides primary on securing the Sea Lanes of Communication (SLOC) for a boost in maritime economic activities, trade and energy security. In the present era, India's economic growth is critically premised on the Indo-Pacific region especially in terms of growth potential and resource dependence. Around 95 percent of the country's trade by volume and 68 percent by value is moved through the oceans.[64] From the lens of Kautaliya[65], the cooperation and collaboration are its tools for maintaining security and prosperity in the region through concerns of global commons & interdependent maritime routes. Thus, the immediate neighbourhood in the maritime domain is considered as associates rather than adversaries.[66] The Indo-Pacific presents itself as a new domain in India's foreign policy engagements which needs a shift in its strategic environment from its continental borders to the maritime space and addressing concerns on key choke points and major sea routes. [67] (Refer Figure 6)

**Figure. 6** : Key Choke points in the Indian ocean.

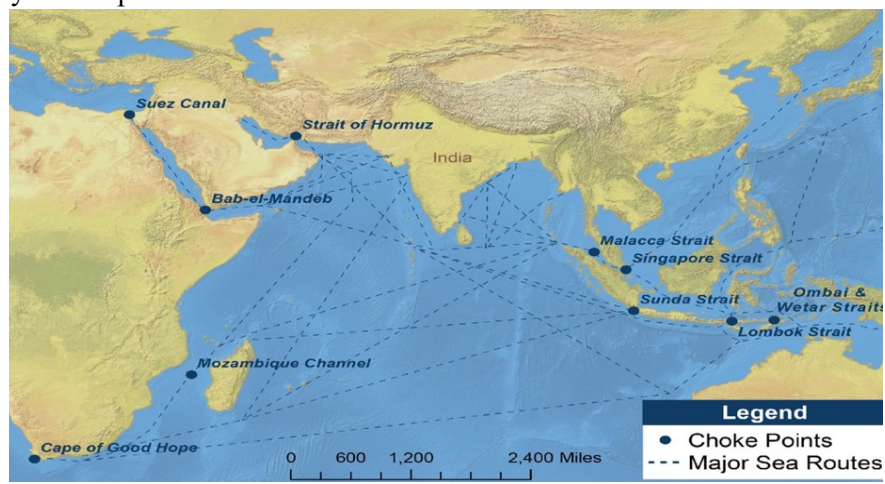

**Source:** Darshana M Baruah, "India in Indo-Pacific: New Delhi's Theatre of opportunity, June 30, 2020, *Carnegie Endowment For International Peace.(Ibid,79.)*

The US-India joint strategic vision for 'Asia Pacific' & Indian Ocean Region provided a boost for regional growth and support for sustainable development & improved connectivity. In August 2021, while chairing the virtual Security Council session and upholding the paramountcy of United nations Convention on the Law of Seas (UNCLOS), Prime Minister Modi highlighted the role of India as a "net security provider" referring its capability in anti-piracy patrolling, maritime security capability, enhancing maritime domain awareness & assistance for natural disasters in the region.[68] India views Indo Pacific as a platform to work with countries having similar interests in manage a rules-based regional order in a cooperative manner and prevent any one power to dominate the region or its SLOC.[69]

---

[63] Kajari Kamal and Gokul Sahni, "India in the Indo-Pacific: A Kautalyan Strategy for the Maritime Mandala," ORF Issue Brief No. 522, February 2022, *Observer Research Foundation.* assessed on 21 June 2022 at https://www.orfonline.org/research/india-in-the-indo-pacific/
[64] Annual Report 2020-2021, Ministry of Ports, Shipping and Waterways, Government of India, p.6. assessed on 24 June 2022 at https://shipmin.gov.in/sites/default/files/AnnualReport2021_0.pdf
[65] Kautaliya was an ancient Indian polymath who was a philosopher and strategist. He is considered as a pioneer in political science & economics and wrote the ancient political treatise "The Arthashastra"
[66] Ibid.
[67] Darshana M Baruah, "India in Indo-Pacific: New Delhi's Theatre of opportunity, June 30, 2020, *Carnegie Endowment For International Peace* assessed on 03 july 2022 at
"https://carnegieendowment.org/2020/06/30/india-in-indo-pacific-new-delhi-s-theater-of-opportunity-pub-82205

## 2.4. Genesis of the QUAD and the present status

The Quadrilateral Security dialogue (QUAD) began as a loose partnership after the Tsunami disaster in 2004 attracting a joint effort for humanitarian and disaster assistance by like-minded countries of US, Australia, Japan and India.[70] However, for next three years the alliance did not see any active participation or activity by any member states. In 2007, Late Japanese PM Shinzo Abe took an initiative to gather the countries together again in the backdrop of the ASEAN regional forum in Philippines, but under unclear objectives and Chinese pressure, the countries hesitated to align on common interests. The four member countries with 1.8 billion people represent a quarter of the world's population and have a combined GDP of over USD 30 trillion. Trade between the QUAD countries exceeded USD440 billion in 2018, with approximately USD 6 trillion in trade with the rest of the countries in the world.[71] The country profile in terms of population, armed forces, GDP and defence budget is shown in Table 1.

**Table 1:** Country profile of QUAD countries

| Country | Population (billion) | Armed Forces (Thousands) | GDP (Trillion in USD) | Defence Budget (Billion in USD) |
|---|---|---|---|---|
| China | 14 | 2035 | 13.6 | 252 |
| India | 13.7 | 1456 | 2.7 | 72.9 |
| US | 328 | 1380 | 20.5 | 778 |
| Japan | 126 | 396 | 5.0 | 49.1 |
| Australia | 25 | 57 | 1.4 | 26.3 |

**Source:** Jeff Smith, "The QUAD 2.0: A Foundation for a Free and Open Indo–Pacific", Heritage Foundation (https://www.heritage.org/global-politics/report/the-QUAD-20-foundation-free-and-open-indo-pacific & Military Spending by country 2022 (https://worldpopulationreview.com/country-rankings/military-spending-by-country)

In 2008, Australia withdrew from the QUAD forum in the wake of pressure from the Chinese government and heightened conflict between the Washington and Beijing. However, in 2012, with augmented military cooperation between US and Australia, Canberra reaffirmed its QUAD commitment by participating in the QUAD naval exercises. The revival of the present QUAD was done in 2017 to counter increased Chinese assertions in the Indo Pacific & a special emphasis was given by Trump and Biden leadership who saw QUAD as a pivot for focusing its interest in the Indo pacific. The QUAD countries exchange views on contemporary global issues such as maritime security, connectivity & infrastructure, critical & emerging technologies, pandemic, humanitarian assistance, disaster relief, cyber security,

---

[68] Anirban Bhaumik, "India asserts its 'net security provider' role in Indian Ocean", August 9, 2021, *The Deccan herald*, assessed on 09 July 2022 at https://www.deccanherald.com/national/india-asserts-its-net-security-provider-role-in-indian-ocean-1017935.html

[69] Premesha Saha, "*India's role in the emerging Dynamics of the indo- pacific*", January 26, 2022 Observer Research Foundation assessed on 3 July 2022 at https://www.orfonline.org/expert-speak/indias-role-in-the-emerging-dynamics-of-the-indo-pacific/

[70] "The QUAD : the origins if the Quadrilateral Dialogue", May 25,2020 , *The Week* assessed on 17 June 2022 at https://www.theweek.co.uk/news/world-news/asia-pacific/956856/the-QUAD-the-origins-of-the-QUADrilateral-security-dialogue

[71] Dr Mukesh Aghi, "*A Free trade Agreement between QUAD nations: Vision or reality*", August 26,2020, *Financial Express*, assessed on 10 July 2022 at https://www.financialexpress.com/economy/a-free-trade-agreement-between-QUAD-nations-vision-or-reality/2066170/

climate change, & education.[72] The first formal meeting of the QUAD was held in March 2021 hosted by Biden administration taking the lead to align the QUAD interests with the US Indo Pacific Policy in the region. Three working groups on vaccines, climate and critical and emerging technologies were formalized. The goal of the QUAD countries was to keep the military and political influence out from the SLOC, maintain a rule based order for freedom of navigation & liberal trading and most importantly create an alternative to "Debt financing" in the region[73]. In the meeting held in September 2021, the leaders exchanged views on shared interest, values and underlying principles on which the QUAD framework was anchored. The leaders of QUAD shared their perspective on situation in Afghanistan & South Asian region. Working groups on Infrastructure coordination, Cyber security and Space coordination were launched to bolster regional infrastructure and exchange satellite data for monitoring climate change, disaster preparedness and responding to other challenges.[74] (Refer Figure 7)

**Figure 7:** QUAD Working Groups and Initiatives as of May 2022

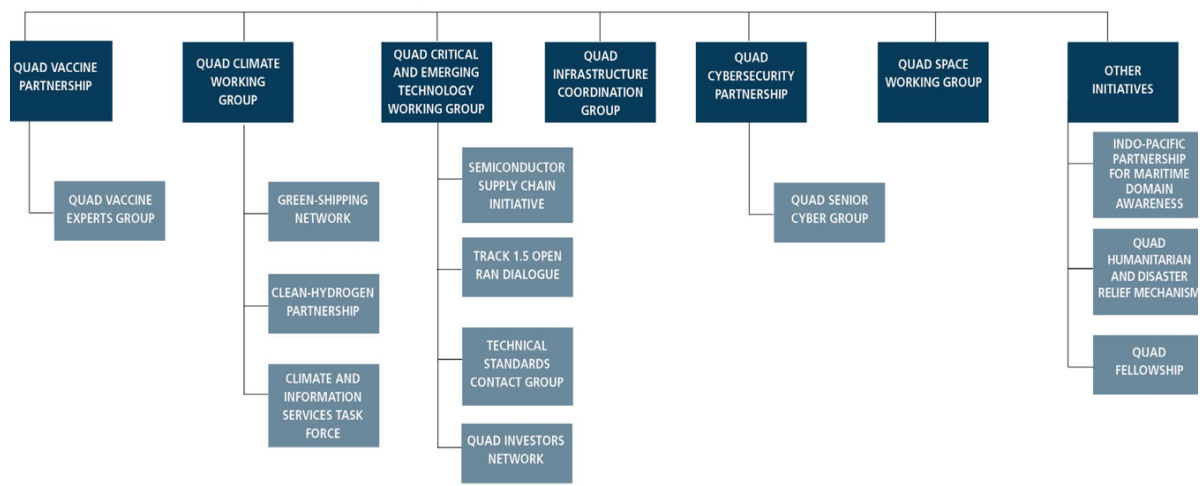

**Source**: *Garima Mohan and Krishti Govella, " The Future of the QUAD and the Emerging Architecture in the Indo pacific, June 2022, The German Marshall Fund of the United States.(Figure1)*

The latest QUAD meeting was held by the leaders of the member nations on May 24, 2022 in Tokyo. The summit which was 4th interaction since its virtual meeting in 2022 witnessed the launch of a new initiatives for continuous collaboration in maritime domain, health, space, climate change, and cyber-security. The members committed themselves to overcome challenges to secure rules-based maritime order especially in the East and South China Seas, and firmly opposed any provocative, coercive, or unilateral actions that seek to alter the status quo.[75] (A summary of the meetings since 2017 is shown in Table 2.)

---

[72] "What is QUAD?", *Business Standard,* assessed on 17 June 2022 at https://www.business-standard.com/about/what-is-QUAD#collapse
[73] Ibid,38.
[74] Prime Ministers' participation in the QUAD Leader's Summit, September 24, 2021, Ministry of External Affairs , Government of India assessed on 24 June 2022 at https://www.mea.gov.in/press-releases.htm?dtl/34324/Prime_Ministers_participation_in_the_QUAD_Leaders_Summit
[75] Ibid,39

**Table 2:** QUAD Working Groups and Initiatives as of May 2022

| Leaders' Summit | Foreign Ministers meetings | Senior officials meeting |
|---|---|---|
| Inaugural virtual Summit, March 12, 2021 | First meeting in new York, September 26, 2019 | Nine meetings since November 2017 to address strategic assessment and progress on cooperation |
| In person Summit in Washington D.C | Second meeting in Tokyo, October 6, 2020 | Intelligence Chief from the countries met in September 2021 during QUAD Strategic Intelligence forum |
| Virtual Summit to discuss the Ukraine conflict, March 4, 2022 | Third meeting Virtual, February 18, 2021 | |
| In person summit in Tokyo, May 24, 2022 | Fourth meeting in Melbourne, February 11, 2022 | |
| **Working Group Meetings**: The six QUAD Working groups have held meetings led by sherpas and sous-sherpas in all the four countries ||| 

**Source**: Garima Mohan and Krishti Govella, *" The Future of the QUAD and the Emerging Architecture in the Indo pacific"*, June 2022, *The German Marshall Fund of the United States.*(Box 1) (Ibid,9.)

With the increased frequency of interaction and commitment by member states, leaders of QUAD have become more aligned by a shared concerns on China's increasingly assertive behaviour in the region and are more willing to define a productive agenda of cooperation which is mutually beneficial.[76] As the QUAD leaders embark on a journey of close cooperation, one of the most important areas being addressed is the maritime security initiative (Indo Pacific Partnership for Maritime Domain Awareness) that will permit member countries to monitor and assess illegal fishing, track dark shipping and perform tactical level activities.[77] A major benefit of this initiative is that it will ensure round the clock virtual presence and improve ability to respond to disaster relief and protection of fisheries.

QUAD has a tremendous potential in the space domain. US has dominated the space historically and now a key indicator of its focus in the space domain is visible with the creation of USSPACECOM and US Space force. In addition to the above, civilian projects which include robotic and human space exploration, human space missions, cislunar system and the lunar exploratory program are being undertaken to maintain its space supremacy. India also has a mature space program which provides a robust, self-reliant and cost effective launch services to foreign countries.[78] Its major achievement include the successful conduct of ASAT test in 2019 and establishing the Defence Space Agency which has an astropolitical and strategic significance in the regional context.[79] Japan established the Japanese Aerospace Exploration Agency (JAXA)[80] in 2003 which serves as a core agency for overall aerospace development and utilization. To augment the military space capability, it is in the process of creating a Domain Mission unit for protecting space assets and will coordinate with USSPACECOM and Space Force. The Defence of Japan 2021 White Paper mentions a special focus in the cyber and space domain.[81] In contrast to all others, Australia is relatively a new player in the space domain. The Australian Space Agency was created in 2018 before which it was dependent on the US for all contributions. In December 2021, the Australian government released a roadmap to develop its space industry with 38 recommendations for space infrastructure and services providing a strategic outline to growth of Australian space communities having ramifications beyond the space sector. [82]

---
[76] Sheila A. Smith, "The QUAD in the Indo-Pacific: What to Know, May 27,2021 Council on Foreign relations assessed on 17 June 2022 at https://www.cfr.org/in-brief/QUAD-indo-pacific-what-know

[77] "QUAD set to launch maritime security initiative to track dark shipping in Indo-Pacific" ,May 24,2022, *The Times of India;* assessed on 17 June 2022 at https://timesofindia.indiatimes.com/world/us/QUAD-set-to-launch-maritime-security-initiative-to-track-dark-shipping-in-indo-pacific/articleshow/91758120.cms

[78] Maj. Gen PK Mallick, '*India in Space Domain: Pathbreaking Developments*', VIF Brief, Vivekananda International Foundation, New Delhi, assessed on 22 June 2022 at
https://www.vifindia.org/sites/default/files/India-in-Space-Domain-Pathbreaking-Developments.pdf

[79] Avantika Menon, "*Space 2.0:Prospects for QUAD Collaboration*", January 2022, VIF Brief, Vivekananda International Foundation, New Delhi assessed on 22 June 2022 at
https://www.vifindia.org/sites/default/files/Space-2-0-Prospects-for-QUAD-Collaboration.pdf

[80] "*Introduction to JAXA*", Japan Aerospace Exploration Agency, assessed on 04 July 2022 at https://global.jaxa.jp/about/jaxa/index.html

### 2.5. Chinese Strategy in the Indo pacific

Chinese grand strategy is wielded through its 2013 BRI program & recognises it as a medium to reshape the regional and international order by building extensive infrastructure projects from Eurasia to the African continent. From the geopolitical lens, Beijing's BRI strategy and the Washington's Indo-Pacific strategy constitute a strategic competition between sea and land powers.[83] China has countered the assessment of US experts by proposing a concept of "community of shared mankind" by promoting the idea to bring tangible economic benefits to the countries in the Indo Pacific.[84] Chinese foreign Ministry statement of May 22 referred to the American strategy being under the banner of 'freedom and openness,' which is keen to "gang up" & create 'small circles'" in an attempt to contain China.[85] Beijing's theatre strategy in the region constitutes the 'String of Pearls'[86] by which it plans to control the strategic choke points in the Indian Ocean region. As part of its expansionary move, Beijing established its first overseas military base 2017 in Djibouti to make its presence felt in the region. It refers to its base in Djibouti as a logistics facility, however, it has stationed PLAN Marines equipped with armoured vehicles and artillery assets.[87] The region as a whole and in particular borders of the Indo-Pacific encompass the maritime segment of the BRI that holds importance for visualized 21st Century Maritime Silk Road.[88] By establishing overseas bases Beijing seeks to create a global logistics infrastructure for expansion & projecting military power in the garb of commercial arrangements to achieve its geopolitical goals.[89] A similar action can be seen in Chinese military arrangements with Cambodia, which undermines the peaceful settlement of disputes, maritime security and freedom of navigation in the ASEAN region. In the South China Sea, PRC claims resources by undermining sovereign rights of South Asian countries and in turn threatening the shipping lanes. It has not offered any coherent explanation or legal basis for the 'Nine dashed

---

[81] *Defence of Japan 2021*, Ministry of Defence, assessed on 04 July at
https://www.mod.go.jp/en/publ/w_paper/wp2021/DOJ2021_EN_Full.pdf
[82] Clare Fletcher, "*The roadmap to the future of the Australian space sector*", December 08, 2021, *Space Australia*, assessed on 24 June 22 at https://spaceaustralia.com/index.php/news/roadmap-future-austra-lian-space-sector.
[83] Daisuke Akimoto, "*China's Grand Strategy and the emergence of Indo-Pacific alignments*", April 14,2021, *Institute for Security & Development Policy*, assessed on 09 July 2022 at https://isdp.eu/chinas-grand-strategy-and-the-emergence-of-indo-pacific-alignments/
[84] Denisov, Igor, Oleg Paramonov, Ekaterina Arapova, and Ivan Safranchuk. "Russia, China, and the Concept of Indo-Pacific." *Journal of Eurasian Studies* 12, no. 1 (January 2021): 72–85. https://doi.org/10.1177/1879366521999899.
[85] "China Foreign minister says USA's Indo-Pacific strategy' doomed", May 22,2022 , *Business Standard* assessed on 16 June 22 at https://www.business-standard.com/article/international/china-foreign-minister-says-usa-s-indo-pacific-strategy-doomed-to-fail-122052200790_1.html
[86] String of Pearls is a geopolitical hypothesis which refers to Chinese commercial infrastructure along the Sea Lanes of Communication. It extends from mainland China to-Strait of Mandeb-Malacca-Strait of Hurmuz and maritime centers in Bangladesh, Sri Lanka, Maldives, Pakistan and Somalia.
[87] Jean-Pierre Cabestan, "China's Djibouti naval base increasing its power", , may 16,2020, *East Asia Forum* assessed on 10 July 2022 athttps://www.eastasiaforum.org/2020/05/16/chinas-djibouti-naval-base-increasing-its-power/
[88] Ibid, 84.
[89] "*Chinese Military Aggression in the Indo-Pacific*", US Department of State, assessed on 18 June 2022 at https://2017-2021.state.gov/chinas-military-aggression-in-the-indo-pacific-region/index.html

Line'[90] claim which is visualized as unlawful by the arbitral Tribunal constituted under the 1982 Law of Sea convention.[91] In the East China sea, PRC strategically deploys its coast guard and military to control and exercise virtual control of the commercial fleet with an aim to intimidate its neighbours. The interference in the Senkaku islands which has been under Japan's administration since the 1971 under the Okinawa reversion agreement are being challenged by the PLAN in various occasions. PRC's open rhetoric on declaration of the "East China Sea Air Defence identification Zone"[92] makes clear about the intentions of the country in the region. PRC also continues to intimidate and threaten the sovereignty of Taiwan. In an attempt to coerce Taiwan into submitting to China, unprecedented air incursions and exercises simulating attacks are taken by PRC threatening regional stability. It plans to use stepping "stone strategy" by threatening to invade the Kinmen, Matsu & Penghu islands in the close vicinity and Pratas atoll in the Spartly chain in South China Sea. The heightened activity in the South China Sea is a well thought of strategy by China to impose a Custom Quarantine to reinforce its territorial claims.[93] Table 3 summarizes Beijing's goals related to activities in South China Sea & the types of actions it takes to support its goals.

**Table 3.** China's Apparent Goals and Supporting Actions for South China Sea

| Ser No | Supporting Actions | Rally support domestically | Deter US | Intimidate neighbors and encourage appeasement/compliance |
|---|---|---|---|---|
| 1 | PLA operations [a] | X | X | X |
| 2 | China Coast Guard operations [b] | X | X | X |
| 3 | Maritime militia swarming | | | X |
| 4 | Dredging fleet and island construction team operation [c] | X | X | X |

a. Includes military exercises, weapons tests, port visits, patrols throughout the SCS, military parades, and participation in echelon formation.
b. Includes deployment of large vessels and participation in echelon formation.
c. Includes large-scale dredging and island building, and construction of permanent facilities on disputed features

**Source**: "U.S.-China Strategic Competition in South and East China Seas: Background and Issues for Congress" January 26, 2022, Congressional Research Service, *(https://crsreports.congress.gov/product/pdf/R/R42784)*

Chinese and Indian troops who share one of the largest land border's in the Indo Pacific region have confronted each other on many occasions in the past decade. In the aftermath of border clashes like Doklam[94] (Standoff at the tri junction with Bhutan in 2017) and the Galwan[95] (June 2020 incident of violent clash in Ladakh leading to fatal casualties at both sides) incidents, both refused to withdraw from its positions occupied and till date the relations remains sour.

---

[90] Nine dash line is a set contested segments claimed by PRC and Taiwan in the South China Sea.
[91] Ibid,47.
[92] East China Sea Air Defense Zone covers most of the East China Sea and overlaps with the airspace in Japanese claimed Senkaku island, South Korea claimed Socotra Rock and Taiwan claimed Diayo islands
[93] *"Control without invasion: Other actions China could take against Taiwan"*, June 17, 2022 *The Times of India,* assessed on 26 June 2022 at https://timesofindia.indiatimes.com/world/rest-of-world/control-without-invasion-other-actions-china-could-take-against-taiwan/articleshow/92271821.cms
[94] Raj Chengappa, "India-China standoff: All you need to know about Doklam dispute", July 17,2017, *India Today*, assessed on 24 June 2022 at https://www.indiatoday.in/magazine/cover-story/story/20170717-india-china-bhutan-border-dispute-doklam-beijing-siliguri-corridor-1022690-2017-07-07
[95] *"The Galwan valley face-off explained through 17 new reports"*, June 22,2020, *Hindustan Times,* assessed on 10 July 2022 at https://www.hindustantimes.com/india-news/the-galwan-valley-face-off-explained-through-17-news-reports/story-UzyJFQdaOEDgJZH2dE1MON.html

**2.5.1 Military capability of China & Likely manifestation in the Indo pacific**

The Chinese Army embarked on a journey of military modernization after the Gulf war drawing lessons for the modern day conflict. With an aim to transform into a modern military PRC undertook radical reforms[96] to prepare itself for projecting beyond the borders. China's defence expenditure, at US$252 billion (2020), is the 2$^{nd}$ largest in the world and is approximately one-third of the US's defence spending.[97] This increase in defence spending is in sync with its long term strategy of becoming a world leader. It has built both offensive and defensive capability to secure the SLOC which are part of the grand Maritime Silk Route project. As part of building a military might and expanding its navy in the blue waters, it is focused to increase the number of nuclear powered submarines, aircraft carriers, ballistic missile launch submarines, and ballistic missile defence to solidify its security concerns in the Indo pacific region.[98] The militarization of the islands in South China Sea has seen the deployment of nuclear powered submarines (Shang I-class), ballistic missile launch submarines (Jin class SSBNs), aircraft carriers (Liaoning & Shandong, Intercontinental ballistic missiles (DF-26), Medium range ballistic missiles (DF-17)). Surface combat ships(Type-054A frigates, Type-056 corvettes & Type-052D destroyers), and combat aircrafts.[99] PLA views Cyber domain as an important pillar of economic and social development. It carries out its cyber operations through "The Ministry of State Security (MSS)" & "Strategic Support Force (SSF)"[100] to through enhanced cyber defence, cyberspace situational awareness, participating in international cyber cooperation, ensuring national information security and maintain national security & social stability.[101][102] SSF emerged from the PLA reorganisation in 2015 and is an umbrella force which integrates electronic warfare, information warfare and cyber operations to the Integrated Theatre Commands. The Chinese cyberwarfare allows multiple agencies and individuals to work towards disrupting adversary's computer networks to paralyse their decision-making capability during peacetime activities and during hostilities.[103] Given the above backdrop, the likely manifestation of China in the Indo pacific is hinged on the following two ways as covered below.

*a) Grey Zone Warfare to advance its interests for regional dominance.* China employs innovative grey zone tactics in the South China Sea, East China Sea and borders with India to gain persistent indefatigable advantage. Table 4. provides a graphical representation of Chinese grey zone tactics against Taiwan, Vietnam, Philippines, India and Japan.

---

[96] Lindsay Maizland, "*Chinese Modernizing Military*", February 5, 2020, Council on Foreign relations, assessed on 10 July 2022 at https://www.cfr.org/backgrounder/chinas-modernizing-military

[97] Air Marshal Amit Dev, *"China's Rise and the Implications for the Indo Pacific"*, April 27, 2022 assessed on 23 June 2022 at https://www.orfonline.org/expert-speak/chinas-rise-and-the-implications-for-the-indo-pacific/

[98] Thangavel K. Balasubramaniam et al., "*China's Rising Missile and Naval Capabilities in the Indo-Pacific Region: Security Implications for India and Its Allies*", June 8, 2022, *Air University* assessed on 17 June 2022 at https://www.airuniversity.af.edu/JIPA/Display/Article/2210972/chinas-rising-missile-and-naval-capabilities-in-the-indo-pacific-region-securit/#sdendnote10sym

[99] Ibid.

[100] Costello, John and Joe, McReynolds. 2018. *China's Strategic Support Force: A Force for a New Era*. Washington, D.C.: National Defense University Press.

[101] "*Chinese Military Strategy*", May 27, 2015, *The State Council Information Office of the People's Republic of China*, assessed on 03 July 2022 at http://english.www.gov.cn/archive/white_paper/2015/05/27/content_281475115610833.htm

[102] Aadil Brar, "*China's Cyber warfare has grown on the back of civilian recruits*", September 22, 2021, *The Print*, assessed on 10 July 2022 at https://theprint.in/opinion/eye-on-china/chinas-cyber-warfare-has-grown-on-the-back-of-civilian-recruits/737651/

[103] "*Taiwan says it faces 5 mn cyberattacks daily: A look at China's cyber capabilities, what it means for India*", November 12, 2021, *First Post*, assessed on 24 June 2022 at https://www.firstpost.com/india/taiwan-says-it-faces-5-mn-cyberattacks-daily-a-look-at-chinas-cyber-capabilities-what-it-means-for-india-10128621.html

**Figure 8.** Chinese Grey Zone tactics

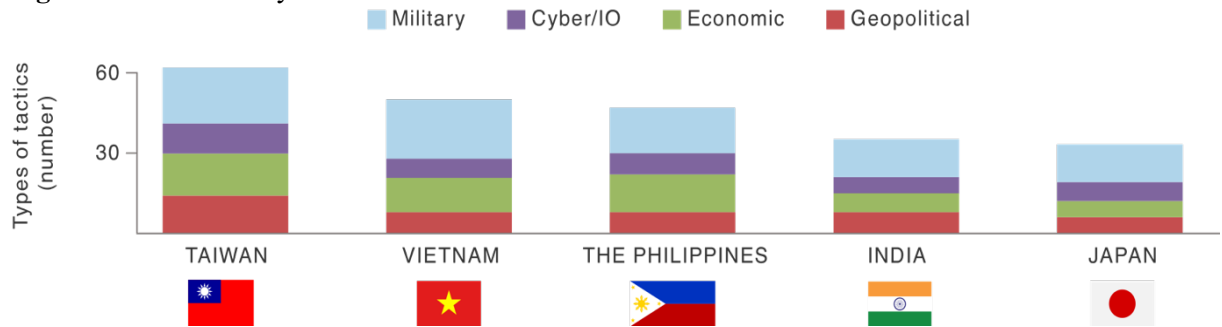

*Source:* Lin, Bonny, et.al., "A New Framework for Understanding and Countering China's Gray Zone Tactics." Santa Monica, CA: RAND Corporation, 2022. (https://www.rand.org/pubs/research_briefs/RBA594-1.html.)

These activities involve deliberately pursuing political objectives via carefully designed operations below the escalatory thresholds of conflict & pursue all non-military and non-kinetic tools such as cyber warfare, legal warfare and information warfare.[104] China may combine grey zone activities with its "Cabbage Strategy" where islands are surrounded layer by layer in a sequence by sending civilian fishing ships in a disputed territory, followed by fisheries patrol vessels, then coast guard ships and consolidation by the PLAN warships.[105] China may heighten the grey zone operations along with multiple surrounding of islands and involve other stakeholders like the armed militia, government owned vessels, PLAF aircrafts and oil rig platforms (which operate with a backup of social media campaigns), GPS & radio interference and Cyber-attacks. As a culmination of the above activities, Beijing would consolidate gains and intend to establish sovereignty over the South China Sea.

*b) Leveraging Maritime Silk Road & Access denial capabilities for domination.* The network of facilities in the Maritime Silk Road is supporting the Chinese commercial domain as it aspires to build an economic chain linking China with countries in Asia, Europe, Middle east and Africa. PLA may further consolidate its formidable access denial capabilities in the near seas (From Japan through Philippines till Malacca) and plan to cover vital choke points between the Indian and Pacific ocean as part of its long range power projection capabilities (Open sea protection). As of 2022, China has 17 overseas ports[106] out of which Djibouti in East Africa, Hambantota in Sri Lanka, Gwadar in Pakistan and a deep water port in Myanmar are important in the context of Indo pacific. China may plan to convert infrastructure to military facilities in no time for power projection. The infrastructure can hold minimum provisions of medical store, rest & recreation facilities, communication detachment and ship repair facility to support a naval combat operation. A likely crisis scenario may emerge wherein the CCP plans for full projection of its force level which entails, mobilization and deployment of PLA forces, enforcement of an Air Defence Identification Zone (ADIZ) and interference with enhanced cyber & space activities.[107] The strategy will involve Active Defence[108] measures to defend littoral border in the Phase-I and develop a blue water navy[109] capable of power projection in Phase-II.

---

[104] Peter Layton, "*Bringing the grey zone into focus*" The interpreter, Lowy Institute, July 22, 2021 assessed on 10 June 2022 at https://www.lowyinstitute.org/the-interpreter/bringing-grey-zone-focus
[105] Ibid.
[106] Saeeduddin Faridi, "China's ports in the Indian Ocean" 19 August, 2021, *Gateway house: Indian Council on Global Relations* assessed on 18 June 2022 at https://www.gatewayhouse.in/chinas-ports-in-the-indian-ocean-region/
[107] Ibid, 98.
[108] Active Defense refers to offensive action at operational & tactical level while retaining defensive posture at strategic level.
[109] A blue water Navy is a maritime force which can operate globally across deep waters.

## 3. Literature Survey

Extensive research has been undertaken by identifying the authors and readings which provide insights into the role of QUAD countries in the geostrategic and security domain, the prospects of QUAD in the near future and the perception of Beijing towards QUAD. Also, a substantial analysis was attempted in two specific fields where authors have proposed ways niche and disruptive technology namely; Artificial Intelligence, Quantum, 5G, Cyber security and Nano technology can be leveraged for use. The literature survey is as covered under.

Bonny Lin covers that Beijing draws a parallel between NATO's expansion in Europe (especially the erstwhile east bloc countries) and US effort in the Indo-pacific to form blocs and alliances to counter China. She argues that the Chinese concern is based on the fact that an increased support by the US to its regional partners and Taiwan may prove to be a similar situation which Russia is facing. She highlights certain lessons Chinese are learning form the Ukraine crisis in terms of dangerous expansion by US military activities which motivates it to be more self-reliant. Her work is admirable as it provides an insight into the likely shifting positions of China in Ukraine and implications for the US. However, it does not cover lessons & options for US in terms of specific initiatives in the space and cyber warfare with the countries of Indo pacific.[110] Sun Chul Jung et.al. postulates that the US Indo Pacific strategy has put the non-Quad Asian states in the strategic dilemma and examines their positions on Chinese rise vis a vis Indo Pacific Strategy.[111] He asserts that the reason for the QUAD not recruiting new members is due to weak commitment to a liberal international order and opposition due to rising China. The authors reasoning in view of the present structure of QUAD is logical, however the author does not touch on the bilateral ties of the non-QUAD countries which may be viewed as a stepping stone to attract these countries for the QUAD membership  Mitika Guha Sarkar in her work claims that Beijing's perception towards the QUAD is complex and systemic. She attributes Beijing's response to QUAD is connected to US' rebuilding Asia strategy beyond military operations, re-emergence of Japan in Asia, Australia's renewed & calculated stand towards China and rise of India. The attribution of the author are apt, however, key issues on the informality of the alliance have not been highlighted which indicates flexible nature of QUAD.[112]

Wooeal Paik et.al. in their work explain the goals of a mini-lateral security cooperation while focusing on the QUAD. The work considers the non-military cooperation focused on infrastructure development as a counter measure to the BRI in the Indo Pacific. The authors claim that China is using a divide and rule strategy to weaken the mini-lateral bonds between the countries by luring them away from the US led efforts. According to them, China is partially fixing its 'Debt trap policy' by reducing the magnitude of its exploitative political and economic interest seeking.[113] The author have highlighted the importance of developing infrastructure as an alternative to the BRI, however, the aspect of dual use of infrastructure for non-military tasks like disaster management have not been stressed.

---

[110] Bonny Lin, *"Implications of the Ukraine Crisis for U.S. Policy in the Indo Pacific"*, May 19, 2022, Center for Strategic & International Studies, assessed on 20 June 2022 at https://www.csis.org/analysis/ukraine-crisis-implications-us-policy-indo-pacific-0
[111] Sung Chul Jung, Jaehyon Lee & Ji-Yong Lee (2021) The Indo-Pacific Strategy and US Alliance Network Expandability: Asian Middle Powers' Positions on Sino-US Geostrategic Competition in Indo-Pacific Region, Journal of Contemporary China,30:127, 53-68, DOI: 10.1080/10670564.2020.1766909
[112] Mrittika Guha Sarkar (2020) *CHINA AND QUAD 2.0: Between response and regional construct, Maritime Affairs*: Journal of the National Maritime Foundation of India,16:1, 110-130, DOI: 10.1080/09733159.2020.1794526
[113] Wooyeal Paik & Jae Jeok Park (2021) *The Quad's Search for Non-Military Roles and China's Strategic Response: Minilateralism, Infrastructure Investment, and Regional Balancing*, Journal of Contemporary China, 30:127, 36-52,DOI: 10.1080/10670564.2020.1766908

Jagannath Panda in his work highlights the importance for India to look for other sustainable non-Chinese frameworks like the Blue Dot network[114]. He argues that it is a geopolitical necessity as it paves way for an alternate global supply chain network. He also opines that it will strengthen the QUAD process and trilateral frameworks between the countries.[115] He emphasizes on the importance of developing an alternative supply chain to counter BRI and stresses on the developing frameworks where China is not present. However, mechanics of bilateral or multilateral framework or informal alliances in the niche technologies having a security dimension have not been highlighted.

Abhijit Singh argues that for the past few years, China has been actively employing the 'Three warfare's' (3Ws) strategy which includes psychological, media, and legal warfare as tools to weaken resolve in South Asia and the Indian Ocean Region. He claims that although the attacks have been confined to Taiwan and South East Asian countries who have a territorial disputes with China, but the strategy is likely to be used against India. He recommends developing institutional frameworks, creating awareness for mitigation, focusing on flexible & durable responses and need for pre-empting attacks as per the shifting political and military situation.[116] The "Active Defence Strategy" and the "Cabbage Strategy" undertaken by Beijing in the South China Sea and East China Sea has not been mentioned in his work. Also, a brief on the information warfare and cyber capabilities would have given a pragmatic view on the way PLA operates.

Paul Kallender-Umezi provides a perspective of Japanese focus on hi-tech infrastructure which according to him provides an added advantage to have collaboration with the partners to build competition against China.[117] Hiroyuki Suzuki argues that, amid the middle power countries like Australia, India, and Indonesia, there are high expectations for Japan to assume an inclusive leadership role in the region owing to its multi-layered relationships in economic, security and deep cultural ties with the countries of Southeast Asia.[118] The authors cover the aspect of Japan taking a lead in infrastructure especially in the ASEAN region, however, the specific domains in which the collaboration can be undertaken by attracting these countries has not been accentuated.

Salvatore Babones points to the model of Malabar exercises to keep peace with aggressive China manoeuvres.[119] He draws a comparison of NATO with QUAD and recommends "promotion of democratic values" and "peaceful resolution of the disputes" as the way forward for QUAD. However, undertaking exercises in the domain of Cyber and Space

---

[114] Blue Dot Network is a multi-stake holder initiative to bring governments, private players and civil society for promoting high quality standards for global infrastructure development

[115] Jagannath Panda, "*India, the Blue Dot network, and the "Quad Plus" Calculus*", *Journal of Indo pacific Affairs,* July 17, 2020, *Air University*, assessed on 17 June 2022 at
https://www.airuniversity.af.edu/JIPA/Display/Article/2278057/india-the-blue-dot-network-and-the-quad-plus-calculus/

[116] Abhijit Singh, "China's Three Warfare's and India", Journal of Defence Studies 7, No 4 (2013), https://idsa.in/system/files/jds_7_4_AbhijitSingh.pdf

[117] Paul kallender-Umezi, "Japan Pursues Rearmament, despite opposition", June 14, 2015, *Defense News,* assessed on 24 June 2022 at https://www.defensenews.com/home/2015/06/14/japan-pursues-rearmament-despite-opposition/

[118] Hiroyuki Suzuki, "*Japan's leadership role in a multipolar Indo Pacific*". October 23,2020,. *Center For Strategic & International Studies CSIS*. United States, 2001, assessed on 19 June 2022 a
https://www.csis.org/analysis/japans-leadership-role-multipolar-indo-pacific

[119] Salvatore Babones, "*The Quad's Malabar Exercises point the Way to Asian NATO*", November 25 2020, *Foreign Policy, assessed on 10 July 2022 at* https://foreignpolicy.com/2020/11/25/india-japan-australia-u-s-quad-alliance-nato/

which have much more significance in the modern warfare has not been covered by the author.

John Lee et.al propose a democracy led tech ecosystem in the Indo Pacific for winning the Geotech battle between the PRC and QUAD countries.[120] The authors highlight creating a competitive alternate to the Chinese led 5G network, establishing a QUAD free & secure digital trade agreement and a QUAD innovation fund. The recommendation by the author provides a vision for creating an alternative to the Chinese dominated techno space and are based on the present underregulating of the use and flow of data. The contribution presents a novel idea of setting up of data governance standards and policy coordination for restricting the economic involvement of China in the manufacturing sector. Integration of Space domain by the QUAD countries via coordinated policies in the field of emerging technologies thereby creating a competitive edge through research, development and operationalizing the space framework has not been addressed by the research study.

Husanjot Chahal et al assesses the existing AI collaboration between the QUAD countries and identifies AI related research strengths which could be leveraged for joint research opportunities like Human computer interaction, data mining, linguistic and theoretical computer science, ML (Machine Learning) & NLP (Natural language processing). The opportunities and path forward presented by the authors are focused toward 1) research which include privacy preserving AI techniques & establishment of a Joint AI research collaboration and 2) investment in the field of AI by incentivizing and growing AI investment as well as coordinating the investment screening & export controls amongst the member countries.[121] Although QUAD offers itself as an alternate to Chinese authoritarian model on technology development and use, but different governance approaches, varying technological capabilities and deep ties in AI research and investment with China pose barriers for creating a collaboration. Specific use cases in the domain of Space and cybersecurity were areas which the authors did not stress upon.

The August 2021, Gateway House report by QUAD economy & Technology task force[122] brought the expert recommendations from industry, politics government and academia in five areas Pharmaceuticals, critical minerals, fintech & cyber security, space & 6G and undersea communication cables. The task force aptly addresses the key areas where an immediate progress on economic and technological issues can be made by the QUAD countries. The study identifies the core element of cybersecurity as reliable hardware and highlights an important aspect of lack of global mechanism to detect cyberattacks and assess privacy risks to the financial system. The recommendations provided by the task force in cyber security domain are on identifying the data protection standards, pushing the technology from the "west" to Indo pacific and on wider accession to the Budapest Convention on cyber security. In the field of 6 G and space domain, the task force proposes space based 6G network where the data transmission would take place through aerospace and space platforms. The technology would include flexible heterogeneous networks with inbuilt encryption for radar & cellular networks and secure green intelligent communication with integrated sensing. The study mentions the lack of an overarching strategy akin to the Chinese "Space information Corridor" as part of the BRI. The task force report is comprehensive and provides recommendations which are of value, however the specific aspects which needs to be addressed as part of the collaboration between the QUAD countries in the space domain over surveillance, communication and remote sensing have not been addressed. Also, although the recommendations on cyber security aspects in the Fintech cover the aspects of protection, but the framework in which the cyber security has to be ensured has not been touched upon.

---

[120] John Lee, Eric Brown, And Thomas Duesterberg, "*Winning the Geo-Tech Battle and Building the QUAD Alliance in the Indo-Pacific*", April 2021, Hudson Institute assessed on 03 July 2022 at https://s3.amazonaws.com/media.hudson.org/Brown%20Lee%20Duesterberg_Winning%20the%20Geo-Tech%20Battle.pdf

[121] Husanjot Chahal, Ngor Luong, Sara Abdulla, Margarita Konaev, "*Assessing AI-related Collaboration between the United States, Australia, India, and Japan*", May 2022 *Center for Security and Emerging Technology* assessed on 02 July 2022 at https://cset.georgetown.edu/wp-content/uploads/QUAD-AI.pdf

[122] "*QUAD Economy & technology Task Force report, A time for concerted effort*", August 2021, Gateway House Report, *Gateway House: Indian Council on Global relations*, assessed on 24 June 2022 at https://www.gatewayhouse.in/wp-content/uploads/2021/08/QUAD-Economy-and-Technology-Task-Force-Report_GH_2021.pdf

Lisa Curtis[123] et.al. provides an assessment on the QUAD activities and policy recommendations with respect to trade, economics, and security & defense across six priority areas which include vaccine, critical & emerging technologies, climate change, infrastructure, space and cybersecurity. In space domain, the authors recommend enhancement of cooperative mechanisms to establish space situational awareness, training to improve interoperability and enhance industry collaboration in the space sector. For Cybersecurity, a shared set of standards, multilateral actions for preventing cyber asset exploitation and civilian workforce interoperability have been recommended. Yet, the key aspects related to space coordination by the respective agencies and cyber security measures required for mitigating the threats related to space infrastructure has not been amplified by the authors.

A. Pereira, B. Biddington, R. Rajagopalan and K Suzuki [124] investigates the perspectives on the emerging geopolitical aspects and its implication for the space activities by leading experts from Australia, India, and Japan. The study brings out that out of all countries, Australia is the smallest investor in space activities, and it exports regime with respect to export of space activities is restrictive but is well located to host ground stations. As a middle power, it can play an important role for stronger space regulation, develop space security matters, space situational awareness and maritime domain awareness. The authors bring out the importance of mission assurance, vulnerability of space assets and space governance. Setting up of 'Space Traffic Management' to prevent accidental and intentional collisions and protecting space-based assets for civil and military purpose are of vital interest. The work highlights the importance of Rendezvous proximity operations (RPO)[125] through transparency & confidence building measures in which the QUAD countries can collaborate on. The author while covering the cyber-attacks on Indian Space Institutions highlights that in view of the shift of Indian Space policy as part of the larger strategic policy shift with the rise of China, there is a requirement for space dialogues for more concerted efforts in developing rules for outer space activities and for counter space capabilities. The point covered by authors on timely access to imagery, strengthened military communication & surveillance during conflict and natural disasters areas is crucial for time sensitive decisions and resource allocation. The authors also agree that the present QUAD summit fact sheet does not identify space domain immersed with the niche technologies which are deemed critical by all four countries.

---

[123] Lisa Curtis, Jacob Stokes, Joshua Fitt, and Andrew Adams, "*Operationalizing the QUAD*" , June 2022, *Center for a New American Security* assessed on 04 July 2022 at
https://s3.amazonaws.com/files.cnas.org/CNAS+Report-IPS-QUAD_Final.pdf

[124] A. Pereira, B. Biddington, R. Rajagopalan and K. Suzuki, "*The QUAD: Implications for Space*," *2021 IEEE Aerospace Conference (50100)*, 2021, pp. 1-15, doi: 10.1109/AERO50100.2021.9438177.

[125] RPO's are needed for a sustained space operations wherein two or more satellites match their plane, altitude and phasing and perform maneuvers to affect their relative states.

Pranay Kotasthane[126] judiciously covers the aspect of 'metacriticality' of semiconductors in the geopolitical & geoeconomics landscape and argues that the QUAD grouping is well placed for collaboration due to the comparative advantage which countries enjoy in the field of semiconductor supply chain. He highlights that the announcement of establishing working group by the QUAD countries in the critical & emerging technologies (C&ET) is a way forward but indicates that there is a lack of consensus which needs to be built. He argues that, although prioritization of specific emerging technologies has been completed, but there is no decision on the mechanics of collaboration. The author recommends multilateral cooperation as a necessity and draws three apt principles 1) thinking security as an ecosystem by forming a QUAD consortium to build semiconductor manufacturing base and cooperate in developing new standards 2) forming a "trust bubble" in semiconductor partnership 3) governments to encourage Research & Development, allow preferential access to access for Exploratory Data Analysis (EDA) tools to QUAD companies and increasing trust in legal enforcement mechanism. The author's work is impressive but lacks specific areas of collaboration especially in the Cyber-Space domain.

Jagannath Panda[127] acknowledges that the emerging technologies such as artificial intelligence (AI), cloud computing, quantum, robotics and blockchain will have a transforming effect in the economy and by 2030 add USD 15 trillion to the global economy. While highlighting Taiwan as a major hub of technology value chain manufacturing and the challenges posed by China, he highlights that the working group can capitalize on advantages of QUAD members with India's data generation strength, Japanese Research and development, Australia's successful public-private partnership and supercomputing capabilities of the US. Although, broad canvas on which the emerging technologies & supply chains needs to be addressed has been recognized, but integration in the space & cyber security domain has not been touched upon.

Avantika Menon[128] provides an authentic account of the new space age 'Space 2.0' and prospects for collaboration by the QUAD countries in the emerging space domain. She provides an insight into the aspect of weaponization in space by covering the aspects of kinetic & non kinetic counter space weapons, electronic counterspace weapons and cyber-attacks on the space systems. She cites Malcolm Davis' assessment that the soft kill weapons are ideal for grey zone operations in the orbit and the threat is real. Her work aptly covers the focus for QUAD collaboration in Space Situational Awareness (SSA), governance and norms in space and development of space solar power stations. The aspects in relation to cybersecurity measures and hardening of ground systems as well as requirement of preparing the space assets for a crisis situation in the Indo pacific did not figure out in her work.

We conclude that considerable research work has been carried out while analyzing the QUAD future to counter Chinese aggressive maneuver. Many authors have proposed future collaboration in harnessing technology by QUAD counties and comprehensively researched the areas of collaboration. However, the research on how the emerging technologies can be enmeshed in the space domain has not been undertaken.

---

[126] Kotasthane, Pranay. (2021). Siliconpolitik: The Case for a QUAD Semiconductor Partnership.
[127] Jagannath panda, "*The QUAD Moves on Critical technologies*", March 19, 2021, *JapanForward* assessed on 09 July 2022 at https://japan-forward.com/asias-next-page-the-QUAD-moves-on-critical-technologies/
[128] Ibid,79.

## 4. Synthesis & Recommendations
### 4.1 Discussion & Opinion

We argue that QUAD countries have a potential of managing the power transition collectively by collaborating on common interests and have a consensus on achieving a higher goal. An alliance which is below the threshold of a formal security alliance but more than a quasi-framework, provides adequate attraction for an alignment that promotes cooperation across common interests without retrenching the sovereignty of members. In the post COVID world order, the ties between the QUAD nations would require greater synergism between the domestic Indo-Pacific vision and the QUAD vision. This collective security concern over Chinese aggressive gestures should aim to strengthen alternate supply chains, present a rule based international order and provide a foundation for an action oriented multilateral institution which covers national interest of all members.[129]

For countries in the Indo Pacific, formal alliances like AUKUS should not be seen as upstaging the QUAD alliance.[130] Both QUAD & AUKUS have an agenda of its own which suits common interests and together should be seen as a pivot for reconfiguring the balance of power in the Indo pacific.

For US, the challenges stemming out of the IPS require strong harmonised efforts as far as the American foreign policy to negotiate with the Indo pacific partners & arrive at a common narrative. The successful implementation of US strategy appears to be delayed due to the ongoing Ukraine crisis as were prior administrations were involved in Middle east and Afghanistan. Washington is likely to face domestic political pressure on its core interest in the region and its rhetoric of liberal democratic institution will go through a litmus test by changed global pattern of Indo pacific partners.[131] India can draw benefits from the Indo pacific Region by having strong relations with the littoral states. Being the largest economy in the region and having a direct security threat from China, it could lead an initiative in the maritime domain through the multilateral forums like the Indian Ocean Rim Association (IORA) and the Indian Ocean Naval Symposium (IONS). It can propose additional agendas for collaboration that cover critical areas in the conventional maritime security, non-traditional security, geographic research & environmental protection.[132] India and Japan have confronted with China over territorial disputes. China's ambassador to Tokyo has publicly criticized Japanese Prime Minister Suga over the claim that, the new QUAD diplomacy demonstrates a "Cold War mentality" and that is "100 percent outdated."[133] Japan under the flagship of FOIP is well poised to engage other states with common interests for regional stability and prosperity. Its strategic partnership with Australia & India with revision of bilateral ties with the US, makes it more aligned towards a collective partnership in the Indo pacific. QUAD partnership is a key pillar of Australian foreign policy and it compliments other bilateral, regional and multilateral cooperation of the country with ASEAN member states and Pacific partners.[134]

---

[129] Poornima Vijaya, "*Australia's Role in the QUAD and Its Crumbling Ties with China*", *Journal of Indo pacific Affairs, Air University*, December 13, 2021 assessed on 17 June 2022 at https://www.airuniversity.af.edu/JIPA/Display/Article/2870644/australias-role-in-the-QUAD-and-its-crumbling-ties-with-china/#sdendnote10sym

[130] "*Aukus has upstaged the QUAD's Indo-Pacific role*", September 22, 2021, *Mint*, assessed on 24 June 2022 at https://www.livemint.com/opinion/online-views/aukus-has-upstaged-the-QUAD-s-indo-pacific-role-11632326428350.html

[131] Ibid,28

[132] Ibid,63.

[133] Sheila A. Smith, "*The QUAD in the Indo-Pacific: What to Know*", May 27, 2021, *Council on Foreign relations* assessed on 117 June 2022 at ,https://www.cfr.org/in-brief/QUAD-indo-pacific-what-know

[134] *QUAD*, Australian Government, Department of Foreign Affairs and Trade, assessed on 22 June 2022, at https://www.dfat.gov.au/international-relations/regional-architecture/QUAD

The Maritime Domain Awareness needs further expansion for dealing with a collective threat in the Indo Pacific region. This would require augmentation of the existing surveillance setup in the region for peace, stability and enhanced cooperation between the QUAD countries & ASEAN in the future. An integrated and fused output of three critical areas namely South East Asia, Indian ocean and pacific islands via fusion centers in India, Singapore, Solomon islands and Vanuatu[135] can provide QUAD countries the "QUAD eyes and ears". Taking into consideration the collective security needs of the QUAD countries and domination of the contested sea lines of communication, it becomes important to leverage the space & cyber security domain together to utilize advanced techniques for utilization in the intelligence gathering. Space Domain Awareness is one area where a credible deterrence needs to be achieved by the QUAD countries. In the recently concluded QUAD summit, the four nations discussed about the semiconductor capacity and the supply chain shortcomings which may pose a major challenge in the coming future. In the Cyber security domain, the responsibilities have been clearly handed over. Australia is in charge of critical infrastructure protection, India is responsible for supply chain resilience and security, Japan administering work force development & talent and US taking lead of software security standards.[136]

China views NATO's expansion and the western liberal order as a subject of controversy and a seed for conflict. Chinese scholars have adopted the term Indo Pacific at regular forums and the think tanks in Beijing have even convened conferences with Indo pacific as the theme & voiced their concerns as an Asian NATO.[137] It sees the extension of support to Taiwan and other partners especially in the Indo pacific as a cause of confrontation and some authors have pointed out that the sanctions which have been posed by the US & the west against Russia were originally tailor made for China in case it invaded Taiwan.[138]

During the Cold War era, US & Soviet union deployed a number of satellites as an umbrella of surveillance to monitor the naval forces.[139] Over the years the space based surveillance has evolved itself as a potent force multiplier with different payloads like high resolution electro-optical imaging satellites, synthetic aperture radar satellites and hyperspectral satellites. These satellites provide near real-time monitoring inputs of the objects in the intended domain and can augment the existing intelligence setup. The emerging model in the space domain allows terrestrial and space systems to be integrated. Space based systems are a strategic resource thus they are not immune to geopolitical conflicts and cyber-attacks on satellites meant to provide services to a particular country could disrupt the infrastructure in another country. Satellite systems originally were not designed with cyber aspects in mind. They have weak encryption and have dependence on legacy systems that cannot be easily patched. There is a need to learn from the lessons of IT networks and agree on standards for developing a common cybersecurity architecture.[140] The recent Ukraine crisis has demonstrated that the space sector will be relevant in the geopolitical conflict and new threat actors will be targeting the space systems to disrupt critical services.[141]

---

[135] Ibid, 62.
[136] Laura Dobberstein, "*QUAD nations pledge deeper collaboration on infosec, data-sharing, and more*", May 25, 2022 assessed on 02 July 2022 at https://www.theregister.com/2022/05/25/QUAD_meeting_tech_agenda/
[137] "*How China views Indo-Pacific*" December 18 ,2020, *The Tribune* , assessed on 22 June 2022 at https://www.tribuneindia.com/news/comment/how-china-views-indo-pacific-13470
[138] Ibid,110.
[139] Cheng, Dean. "The Importance of maritime Domain Awareness for the Indo-Pacific QUAD Countries." Washington, DC: The Heritage Foundation, March 6,2019 assesses on 24 June 2022 at http://thf_media.s3.amazonaws.com/QUAD%20Plus/2019%20Conference%20Papers/Cheng,%20Dean%20-%20The%20Importance%20of%20MDA%20for%20the%20Indo–Pacific%20QUAD%20Countries_JLedit.pdf
[140] Josh Lospinoso, "*Space race needs better cybersecurity*", January 13,2022, *The Hill* assessed on 22 June 2022 at https://thehill.com/opinion/cybersecurity/589542-space-race-needs-better-cybersecurity/

There are number of ways in which hackers can pose a disruptive strategic threat to satellite system. They could compromise the Ground control equipment by spoofing or injecting malware. A study of Global Navigation Satellite System (GNSS)[142] spoofing signals by C4AD identified 9,883 suspected instances across 10 locations that affected 1,311 civilian vessel navigation system–GNSS spoofing events across Russian federation [143] Low cost, ease of deployment & commercial availability of spoofing technologies, empower states and nonstate actors to endanger global navigation safety.[144] China has been suspected of offensively acting in satellite attacks. There were 14 reported incidents including disruption of National Oceanic and Atmospheric Administration network in 2014. Against the QUAD nations, China has pursued an offensive cyber agenda by launching proxies like the nationalistic hacking groups and hired cybercriminals primarily targeting critical infrastructure in the QUAD countries and other nations in the Indo Pacific resulting in the theft of business secrets, intellectual property and sensitive information, and disruption of critical services.[145]

As the satellite operations use open protocols, they offer many vectors for hacking by cyber attackers. Human factor and supply chains are the main vulnerabilities which can be exploited by the adversary. Ground systems are the weak points in the satellite system architecture which encompass TT&Cs & Satellite Operations Centers to Network Operation Centers through teleports, gateways, & earth stations.[146] The continued attacks demands a revamp and hardening of existing satellite based communication architecture and ground based command and control system. It is surprising that neither the space policy nor cybersecurity policy is prepared for the challenges in the Cyber Space Domain.[147] Cyber threats & capabilities continue to proliferate & evolve, so should the QUAD countries especially United States & India develop ability to deflect and counteract against threats to keep satellite protection as a central priority for C4ISR.[148]

The ongoing Ukraine war has made Chinese opinion more strong on its view of the dangerous external environment in view of the US military activities and NATO expansion which could lead to a conflict in the Indo-pacific. The message from Beijing is clear in tone that any alignment of countries in an formal or informal alliance in the Indo pacific will be seen as confrontational. China insists on an architecture which is hierarchical in nature in which political economic and security interest are self-defined for projection of power. Till the time China can be persuaded to accept a multipolar order, Indo pacific strategy will remain in its essential role to counterbalance China with QUAD as its nucleus.[149]

---

[141] Juliana Suess, "*Jamming and Cyber Attacks: How Space is Being targeted in Ukraine*, April 5,2022, *Royal United Services Institute,* assessed on 10 July 2022 at https://www.rusi.org/explore-our-research/publications/commentary/jamming-and-cyber-attacks-how-space-being-targeted-ukraine

[142] GNSS comprises of a constellation of international satellites and the program include Global positioning System (GPS), China's Bediou, Russia's GLONASS and Europe's Galielo.

[143] "*Above us only stars, exposing GPS spoofing in Russia and Syria*", *C4 ADS* , assessed on 23 June 2022 at https://static1.squarespace.com/static/566ef8b4d8af107232d5358a/t/5c99488beb39314c45e782da/1553549492554/Above+Us+Only+Stars.pdf

[144] Jim Edwards, *"The Russians are screwing with the GPS system to send bogus navigation data to thousands of ships, think tank claims*", April 14, 2019, *taskandpurpose assessed on 22 June 2022 at* https://taskandpurpose.com/news/russian-hacking-gps-navigation-gnss/

[145] Sameer Patil, "*Cyber resilience in the QUAD*", October 14, 2021, Gateway House, *Indian Council of global Relations* assessed on 26 June 2022 at https://www.gatewayhouse.in/cyber-resilience-QUAD/

[146] Mark Holmes, "*The Growing Risk Of A Major Satellite Cyber Attack*", *ViaSatellite,* assessed on 22 June 2022 at https://interactive.satellitetoday.com/the-growing-risk-of-a-major-satellite-cyber-attack/

[147] Chuck brooks, "*The Urgency to Cyber-Secure Space Assets*" ,April 22,2022, *Forbes*, assessed on 26 June 2022 at https://www.forbes.com/sites/chuckbrooks/2022/02/27/the-urgency-to-cyber-secure-space-assets/?sh=3c0bed4751b1

[148] "*The Rise of Space-Based C4ISR*" , *Defense One*, assessed on 09 July 2022 at https://www.defenseone.com/insights/cards/space-based-c4isr/7/?oref=d1-cards-cardstack-toc

The Russian invasion of Ukraine witnessed large scale disruption of the communication links through satellite modems and showed us the vulnerabilities in the satellite chain which could be exploited by an adversary.[150] Space & Cyber domains have changed the traditional notions of conflict by expanding its range to a global enterprise, creating potential uncertainty thereby driving the need for technological innovation.[151] As the space domain is now central to modern warfare, there is a need to align the effort for a greater resilience against the emerging threat & eventually contribute to crisis outcomes in space domain using space based systems. The QUAD countries need to fill the vacuum on how cyber hardened satellites spread over large constellations and geographically distributed launch sites can be utilized as a cost-effective solution for improving the space-based surveillance to meet the strategic demands in the Indo Pacific region.

**4.2 Policy recommendations for QUAD countries**

---

[149] Ibid,136.
[150] Gordon Corera, "*Russia hacked Ukrainian satellite communications, officials believe*", March 25,2022, *BBC News,* assessed on 09 July 2022 athttps://www.bbc.com/news/technology-60796079
[151] Jerry Dre, "*Space, Cyber and Changing Notions of War*", August 30, 2017, *Small Wars Journal*, assessed on 24 June 2022 at https://smallwarsjournal.com/jrnl/art/space-cyber-and-changing-notions-of-war

Before attempting to make recommendations about the QUAD alliance, one aspect which merits attention is that the security bargaining architecture and practical alliance proposals have evolved over time. Informality of alliances still offers greater diplomatic and security benefits to nation states at a less political and economic cost.

Table 4. Recommendations for QUAD countries

| Ser No | Recommendation | Sphere of Influence |
|---|---|---|
| 01 | Coordinate response to creation of Infrastructure | Economic Development & promotion of human rights |
| 02 | Research Collaboration in niche technologies | ISR & Cyber and Space Domain |
| 03 | Joint exercises in Space & Cyber Domain | Maritime Security, Crisis management and Strengthen Cyber space domain |
| 04 | Launch low cost regional satellite constellation | Space Domain Awareness |
| 05 | Create Robust Cyber-Space infrastructure | Cyber and Space Domain |

Based on the synthesis and literature survey carried out, certain recommendations for the QUAD countries are as follows:- (Refer Table 4 above)

1. *Coordinated response to creation of joint infrastructure.* To make the alliance more effective there would be a requirement to undertake a coordinated role to provide economic and development assistance, infrastructure development and promotion of human rights in the Indo pacific region. Considering defence and cooperation in infrastructure development as the most plausible avenues for tighter QUAD bonding, working groups in these areas would serve as the foundation for further activity.[152] QUAD should evolve a plan to integrate and put a balanced development in the Indo pacific as the top agenda for consultative engagement. The "Blue Dot Network," announced in 2019 as an initiative of the U.S. Overseas Private Investment Corporation (OPIC), Japan Bank of International Cooperation (JBIC), and the Australian Department of Foreign Affairs and Trade (DFAT) provides a perfect platform to realize effective infrastructure development element of QUAD. Special projects like creation of ports, dockyards. repair facilities, airfields and communication centres should be conceptualized by the QUAD countries to strengthen the interoperability and interdependence between them. These facilities should be made for a dual purpose to cater for peace time activities as well as to sustain forces during wartime.

2. *Research collaboration in niche technologies.* An umbrella of surveillance to jointly monitor the movements of shipping vessels in the Indo Pacific has been announced in the recently concluded QUAD summit in May 2022. This will involve use of commercial satellites to check dark shipping, illegal fishing, smuggling and piracy within the maritime boundaries of the countries. The fusion centres will undertake the collation and analysis of information, it will help in detecting any tactical activities which the PLAN undertakes. A recommended way forward is to take research collaborations in the niche technologies like AI, Quantum technology and cyber security specifically to address problems of communication delays, data processing and prediction of threats. Leveraging the technological research into practical application for operational use like use of AI/ML for contextual analysis in imagery intelligence will help better prediction and decision making. Member countries should also elevate areas of strategically important science & technology collaborations for sustained investment. A functional collaboration with enhanced interoperability and interdependence is required in the areas of ISR (intelligence, surveillance and reconnaissance), maritime security, logistic support (both near & abroad) encompassing civil military defence & industrial sector.

---
[152] Patrick Gerard Buchan and Benjamin Rimland, "*Defining the Diamond: The Past, Present, and Future of the Quadrilateral Security Dialogue",* March 16,2020, *CSIS*, assessed on 04 July 2022 at
 https://www.csis.org/analysis/defining-diamond-past-present-and-future-Quadrilateral-security-dialogue.

*3. Joint Exercises in Cyber and Space domain.* QUAD should aim at accelerating cooperation where already multilateral framework exists especially in the Maritime domain. Two cost-effective ways to do this would be, Firstly to enhance interoperability and jointness through an expanded defence exercises and wargaming in the "cyber and space domain" akin to the Malabar exercises[153] but increased in scope to paint realistic scenarios in these domains. Secondly, specially nominated experts in the Cyber and space domain should undergo joint training on countering cyber-attacks. The exercises and joint training will help in developing joint capability and foster trust & familiarity between countries against Chinese Cyber Warfare.

*4. Launch low cost regional satellite constellation.* Use of low cost miniature satellites constellations complemented by government support in responsive space launch capabilities can increase the foot print of countries to match the growing assertion of China. Australia can offer the launch centre's at Nhulunbuy in Northern Australia being closer to equator & Whaler's bay (South Australia) to support the space launches of QUAD countries. Options for collaboration in looking beyond immediate LEO and near earth GEO orbits, should be undertaken by the QUAD.[154] The manufacturing of the satellites should be undertaken with a regional homegrown electronics ecosystem. Due consideration to the supply chain risk management should be given and all critical components should be developed in a Joint fabrication facility backed by funding from QUAD.

*5. Regional Space Security Working Group.* To preserve the collective interests of the countries and enhanced security framework in space, there is a need to create a 'Space Security Working Group'. The leading space countries like India and US should collaborate and develop a consensus on extending integrated security in the space domain.[155] New generation satellite players are seeking to offset the associated problems due to loss of a single satellite. Satellite miniaturization, laser communication, small satellite constellation when employed could deter a counterattack in space due to high cost of launching a DA (Direct Ascent) ASAT will exceed tactical value of a missile attack for destroying a single satellite. The concept of Operationally Responsive Space (ORS) can be a key stone technology which uses, reusable launch vehicles, satellite prepositioning near launch sites and a multidomain launch effort from land, air, and sea.

*6. Robust Integrated Cyber & Space Infrastructure.* QUAD should recognize the importance of cyber and space as a collective identity for hardening the space systems against cyber-attacks. An infrastructure for information sharing for the security & resiliency of the space systems and an interagency risk management structure should be established. A cyber security framework can be used for profiling that helps to communicate the existing cyber security posture and organize cyber security related tasks. This framework should be used to establish scope & priorities, orient & create a current profile in Phase-I and undertake a thorough risk assessment, target profiling and gaps determination in the Phase-II.[156] Threat intelligence & best practices from engineering and cyber experts of regional industries from manufacturing, launch services & orbit operations should be integrated. Also, for building international consensus on security of space systems, explicit amendments to norms using Article 7 of Outer Space Treaty should be considered by QUAD members.

---

[153] Michael J. Green and Andrew Shearer, "*Countering China's Militarization of the Indo-Pacific*", April 23, 2018 *War on the Rocks* assessed on 02.July 2022 at https://warontherocks.com/2018/04/countering-chinas-militarization-of-the-indo-pacific/
[154] Malcolm Davis, "*The QUAD must go to space*", April 7, 2021 *Australian Strategic Policy Institute*, assessed on 01 July at https://www.aspistrategist.org.au/the-QUAD-must-go-to-space/
[155] Clementine G. Starling, Mark J. Massa, Lt Col Christopher P. Mulder, and Julia T. Siegel, *"The Future of Security in Space: A Thirty-Year US Strategy"* April 2021, *Atlantic Council* assessed on 02 July 2022 at https://www.atlanticcouncil.org/wp-content/uploads/2021/04/TheFutureofSecurityinSpace.pdf
[156] Matthew Scholl & Theresa Suloway, "*Introduction to Cybersecurity for Commercial Satellite Operations*", National Institute of Standards and Technology, February 2022, assessed on 02 July 22 at https://nvlpubs.nist.gov/nistpubs/ir/2022/NIST.IR.8270-draft2.pdf

### 4.2.1 Limitations

Although the research reached its intended aim of recommending policy decisions for the QUAD as an informal alliance especially in the cyber-space domain, there were some limitations to the study undertaken. Firstly, a detailed analysis on the extension of QUAD alliance beyond the Indo Pacific region was not undertaken due to limited scope & paucity of time. Secondly, the deterrence options with the QUAD countries in terms of nuclear strategic stability was not discussed. The growth potential and scope of collaboration has been extensively covered, however due to the divergent threat perception by the QUAD countries in terms of absence of territorial disputes (Except India), varying national priorities and risks of retaliation by China does not make the prospects of states for effective cooperation, a smooth sail. A future research may be undertaken on the aspects of semiconductor supply chain affecting the development of defence systems and prospects of civil defence fusion through QUAD.

### 5. Conclusion

In this paper, we considered the importance of informal alliance's in the present geopolitical environment and examined the motives of QUAD countries & growing Chinese assertiveness in the Indo Pacific region. The opinions brought a clarity on the convergence and need for an arrangement to defend key interests of the member countries due to growing security concerns in the region. An extensive literature survey was carried out with a focus on two critical areas of cyber and space domain to identify certain gaps in the studies carried out. A major observation derived from the survey was that the present studies focus on cyber and space domains exclusively but do not focus on the Cyber-Space collectively as an important field. We furnished key policy recommendations keeping in mind the present challenges which the QUAD countries face geopolitically. Certain limitations while carrying out the study have also been summarized along with prospects of related future work. In the years to come, it is predicted that the QUAD countries will strengthen their stand in the Indo pacific in a more concrete manner. QUAD will have to work out a detailed plan for enhanced security cooperation with ASEAN countries like, South Korea, Vietnam, Indonesia and the Philippines. Although, these countries are less likely to join a collective security framework against China but it is in the interest of the Indo Pacific region to pay greater attention for enhancing economic cooperation with these countries to present an alternative to Chinese investments. US needs to take advantage of the present global focus on Ukraine to strengthen Taiwan's international standing in terms of its sovereignty, defensive capacity and resilience. The fighting capability of Ukraine has proven its worth and has set the tone for countries like Taiwan to protect its own sovereignty through self-defence. QUAD can include Taiwan in the existing framework and expand efforts to strengthen Taiwan's defensive mechanism by providing ready reserves, platforms, network security, energy and transportation along with civil defence planning. Indo Pacific region would certainly be at the center stage to global peace and prosperity in the near future. The Ukraine crisis has demonstrated that the alliances still hold importance as far as the security and regional stability of the countries is concerned and thus, makes QUAD countries more determined in their resolve to counter China.

62. "*Control without invasion: Other actions China could take against Taiwan*" ,June 17, 2022 *The Times of India,* assessed on 26 June 2022 at  https://timesofindia.indiatimes.com/world/rest-of-world/control-without-invasion-other-actions-china-could-take-against-taiwan/articleshow/92271821.cms

63. Raj Chengappa, "India-China standoff: All you need to know about Doklam dispute", July 17,2017, *India Today*, assessed on 24 June 2022 at https://www.indiatoday.in/magazine/cover-story/story/20170717-india-china-bhutan-border-dispute-doklam-beijing-siliguri-corridor-1022690-2017-07-07

64. Thangavel K. Balasubramaniam & Ashok Kumar Murugesan, China's Rising Missile and Naval Capabilities in the Indo-Pacific Region: Security Implications for India and Its Allies, June 8, 2022 assessed on 17 June 2022 at   https://www.airuniversity.af.edu/JIPA/Display/Article/2210972/chinas-rising-missile-and-naval-capabilities-in-the-indo-pacific-region-securit/#sdendnote10sym

65.  "*Chinese Military Strategy*" ,May 27, 2015, , *The State Council Information Office of the People's Republic of China*, assessed on 03 July 2022 at http://english.www.gov.cn/archive/white_paper/2015/05/27/content_281475115610833.htm

66. Aadil Brar, "*China's Cyber warfare has grown on the back of civilian recruits*", September 22, 2021*, The Print*, assessed on 10 July 2022 at https://theprint.in/opinion/eye-on-china/chinas-cyber-warfare-has-grown-on-the-back-of-civilian-recruits/737651/

67. Abhijit Singh, "*China's Three Warfare's and India*", Journal of Defence Studies 7, No 4 (2013), https://idsa.in/system/files/jds_7_4_AbhijitSingh.pdf

68. "*Advantage at Sea- Prevailing with Integrated All-Domain Naval power*", December 2020, assessed on June 17  2022 at https://media.defense.gov/2020/Dec/16/2002553074/-1/-1/0/TRISERVICESTRATEGY.PDF

69. Akshobh Giridharadas, "*Why the AUKUS help the QUAD*", October 21, 2021, Observer Research Foundation assessed on June 15,2022 at https://www.orfonline.org/expert-speak/why-the-aukus-helps-the-QUAD/

70. Air Marshal Amit Dev,  *"China's Rise and the Implications for the Indo Pacific"*, April 27, 2022, *Observer Research Foundation,*  assessed on 23 June 2022 at https://www.orfonline.org/expert-speak/chinas-rise-and-the-implications-for-the-indo-pacific/

71. Anirban Bhaumik, "*India asserts its 'net security provider' role in Indian Ocean*", August 9, 2021, *The Deccan herald*, assessed on 09 July 2022 at https://www.deccanherald.com/national/india-asserts-its-net-security-provider-role-in-indian-ocean-1017935.html

72. Ashish Dangwal, "*In A Massive Policy Change, Japan Permits Export Of Its Fighter Jets, Missiles To 12 Countries Including India – Reports*", May 28, 2022, *The EurAsian Times, assessed on 26 June 2022 at https://eurasiantimes.com/japan-permits-export-its-fighter-jets-missiles-to-12-countries/*

73. A. Pereira, B. Biddington, R. Rajagopalan and K. Suzuki, "*The QUAD: Implications for Space*," *2021 IEEE Aerospace Conference (50100)*, 2021, pp. 1-15, doi: 10.1109/AERO50100.2021.9438177.

74. Lindsay Maizland, "*Chinese Modernizing Military*", February 5, 2020, Council on Foreign relations, assessed on 10 July 2022 at https://www.cfr.org/backgrounder/chinas-modernizing-military

75. "*Taiwan says it faces 5 mn cyberattacks daily: A look at China's cyber capabilities, what it means for India*", November 12, 2021, *First Post*, assessed on 24 June 2022 at https://www.firstpost.com/india/taiwan-says-it-faces-5-mn-cyberattacks-daily-a-look-at-chinas-cyber-capabilities-what-it-means-for-india-10128621.html

76. Sameer Patil, "*Cyber resilience in the QUAD*",  October 14, 2021, Gateway House, *Indian Council of global Relations* assessed on 26 June 2022 at https://www.gatewayhouse.in/cyber-resilience-QUAD/

77. Sheila A. Smith, "*The QUAD in the Indo-Pacific: What to Know*", May 27, 2021, *Council on Foreign relation*s  assessed on 117 June 2022 at https://www.cfr.org/in-brief/QUAD-indo-pacific-what-know
39

## **ABBREVIATIONS**

AI                      -        Artificial Intelligence
ADIZ                -        Air Defence Identification Zone

| | | |
|---|---|---|
| ASEAN | - | Association of South East Asian Countries |
| AUKUS | - | Australia-United Kingdom-United States |
| ANZUS | - | Australia- New Zealand-United States |
| ASAT | - | Anti-Satellite Weapon |
| BRI | - | Belt and Road Initiative |
| CCP | - | Chinese Communist Party |
| C&ET | - | Critical & Emerging Technologies |
| C4ISR | - | Command, Control, Communications, Computers, Intelligence, Surveillance and Reconnaissance |
| DFAT | - | Department of Foreign Affairs and Trade |
| DWP | - | Defence White Paper |
| DSU | - | Defence Strategic Update |
| EDA | - | Exploratory Data Analysis |
| FIGO | - | Formal Intergovernmental Organization |
| FOIP | - | Free and Open Indo Pacific |
| FPWP | - | Foreign Policy White Paper |
| GDP | - | Gross domestic Product |
| GEO | - | Geostationary Equatorial Orbit |
| GNSS | - | Global Navigation Satellite System |
| IGO | - | Intergovernmental Organization |
| IIGO | - | Informal Intergovernmental Organization |
| IPS | - | Indo Pacific Strategy |
| IORA | - | Indian Ocean Rim Association |
| IONS | - | Indian Ocean Naval Symposium |
| JBIC | - | Japan Bank of International Cooperation |
| JAXA | - | Japanese Aerospace Exploration Agency |
| LEO | - | Low Earth Orbit |
| ML | - | Machine Learning |
| MSS | - | Ministry of State Security |
| NLP | - | Natural language processing |
| OPIC | - | Overseas Private Investment Corporation |
| ORS | - | Operationally Responsive Space |
| PLA | - | Peoples Liberation Army |
| PLAAF | - | Peoples Liberation Army Air Force |
| PLAN | - | Peoples Liberation Army Navy |
| PRC | - | Peoples Republic of China |
| QUAD | - | Quadrilateral Security Dialogue |
| RPO | - | Rendezvous Proximity Operations |
| SSA | - | Space Situational Awareness |
| TPP | - | Trans pacific partnership |
| TT&C | - | Telemetry, tracking and Control |
| USSPACECOM. | - | United States Space Command |

**List of Figures**

# List of Tables